\documentclass[a4paper,12pt]{article}
\usepackage{amsmath}

\usepackage{amsfonts}
\usepackage[top=2cm, bottom=2cm, left=2cm, right=2cm]{geometry}
\usepackage{cancel}
\usepackage{slashed}
\usepackage{graphicx}
\def\Xint#1{\mathchoice
{\XXint\displaystyle\textstyle{#1}}%
{\XXint\textstyle\scriptstyle{#1}}%
{\XXint\scriptstyle\scriptscriptstyle{#1}}%
{\XXint\scriptscriptstyle\scriptscriptstyle{#1}}%
\!\int}
\def\XXint#1#2#3{{\setbox0=\hbox{$#1{#2#3}{\int}$ }
\vcenter{\hbox{$#2#3$ }}\kern-.6\wd0}}

\def\dashint{\Xint-}
\numberwithin{equation}{section}
\title{\bf $AdS_3/CFT_2$, finite-gap equations and massless modes}
\author{Thomas Lloyd and Bogdan Stefa\'nski, jr \\ \\ \it Centre for Mathematical Science, City University London,\\ \it Northampton Square, London EC1V 0HB, UK}
\date{}
\begin{document}
\maketitle

\begin{abstract}
It is known that string theory on $AdS_3\times M_7$ backgrounds preserving 16 supercharges is classically integrable. This integrability has been previously used to write down a set of integral equations, known as the finite-gap equations. These equations can be solved for the closed string spectrum of the theory. However, it has been known for some time that the $AdS_3\times M_7$ finite-gap equations do not capture the dynamics of the massless modes of the closed string theory. In this paper we re-examine the derivation of the $AdS_3\times M_7$ finite-gap system. We find that the conditions that had previously been imposed on these integral equations in order to implement the Virasoro constraints are too strict, and are in fact not required. We identify the correct implementation of the Virasoro constraints on finite-gap equations and show that this new, less restrictive condition captures the complete closed string spectrum on $AdS_3\times M_7$.
\end{abstract}
\newpage
\section{Introduction}
The integrability approach to the gauge/string correspondence has provided strong evidence for the duality between
certain strongly coupled gauge theories and their gravitational string duals. For a review and a complete list of references see \cite{Beisert:2010jr}. The principal success of this approach has been the maximally supersymmetric dual pair of ${\cal N}=4$ super Yang-Mills theory (SYM) and Type IIB string theory on $AdS_5\times S^5$, which has 32 real supercharges (see for example~\cite{Minahan:2002ve}).\footnote{Integrability methods can be extended to orbifolds, orientifolds and deformations of this dual pair. See for example \cite{Zoubos:2010kh}.}  Following the discovery of 2+1-dimensional super Chern-Simons theories with a large amount of supersymmetry \cite{Bagger:2006sk,Bagger:2007jr,Gustavsson:2007vu,Aharony:2008ug} and their gravitational duals \cite{Aharony:2008ug}, the integrability approach was extended to ${\cal N}=6$ ABJM theory and its gravitational dual the Type IIA string theory on $AdS_4\times\mathbb{CP}^3$, see for example~\cite{Minahan:2008hf,Arutyunov:2008if,Stefanski:2008ik,Gomis:2008jt,Gromov:2008qe}. This dual pair has 24 real supercharges. It was found that many of the integrability methods employed in the study of the maximally supersymmetric $AdS_5/CFT_4$ example could easily be extended and adapted to the $AdS_4/CFT_3$ case. One novelty of the $AdS_4/CFT_3$ dual pair is the presence in the spectrum of the string theory of excitations of different masses. This is easiest to see in the plane-wave limit of the theory \cite{Blau:2002dy,Berenstein:2002jq}, where we see that there are `light' states of mass $\frac{1}{2}$ and `heavy' states of mass 1. These two types of excitations enter the integrability machinery in a different way to one another. The `light' states can be thought of as elementary particles in the spin-chain description while the `heavy` states appear from the spectrum of these elementary particles.

The integrability approach has more recently been applied to the $AdS_3/CFT_2$ correspondence \cite{Babichenko:2009dk}. The $AdS_3/CFT_2$ dual pairs have at most 16 supersymmetries and there are two classes of string geometries with 16 supercharges: $AdS_3\times S^3\times T^4$ and $AdS_3\times S^3\times S^3\times S^1$.\footnote{The backgrounds $AdS_3\times S^3\times K3$ for the purpose of this paper can be simply thought of as orbifolds of $AdS_3\times S^3\times T^4$.}\footnote{Throughout this paper we restrict our attention to these cases of the $AdS_3/CFT_2$ correspondence with Ramond-Ramond (R-R) background. The mixed Neveu Schwarz-Neveu Schwarz (NS-NS) R-R flux background for $AdS_3/CFT_2$ was also shown to be integrable in~\cite{Cagnazzo:2012se}. Since then there has been much progress in understanding the integrability properties of these backgrounds~\cite{Hoare:2013pma,Hoare:2013ida,Hoare:2013lja}.} In these spacetimes the radii of the $AdS_3$ and $S^3$ spaces are related to one another. For $AdS_3\times S^3\times T^4$ one has
\begin{equation}
R_{AdS_3}=R_{S_3}\ ,
\end{equation}
while for $AdS_3\times S^3\times S^3\times S^1$ one has
\begin{equation}
\frac{1}{R_+^2}+\frac{1}{R_-^2}=\frac{1}{R^2}\ ,
\end{equation}
where $R_\pm$ are the radii of the two 3-spheres and $R$ is the $AdS_3$ radius. This latter relationship leads one to define
\begin{equation}
\cos^2\phi\equiv\frac{R^2}{R_+^2}\ .
\end{equation}
The moduli of $T^4$ and $S^1$ are free parameters of the dual pairs. The presence of this moduli space (when combined also with S-duality) is one of the major novel feature of the $AdS_3/CFT_2$ correspondence as compared with its higher-dimensional higher-supersymmetric cousins. Another important difference is the presence of {\em massless} as well as massive excitations. In the plane-wave limit of $AdS_3\times S^3\times T^4$ one finds states with $m=0$ and $m=1$, while the plane-wave limit of $AdS_3\times S^3\times S^3\times S^1$ has states of mass $m=0,\sin^2\phi,\cos^2\phi$ and $m=1$.\footnote{The massless modes in the $AdS_3\times S^3\times T^4$ theory come from the $T^4$ bosons and their superpartners. In the $AdS_3\times S^3\times S^3\times S^1$ theory one of the massless bosons comes from the $S^1$ direction, while the other comes from the fact that, in choosing the light-like geodesic needed for the plane-wave limit, there is freedom in which linear combination of geodesics on the two $S^3$ factors one picks. The plane-wave limits of the $AdS_3$ backgrounds were investigated in~\cite{Lu:2002kw,Hikida:2002in,Gomis:2002qi,Gava:2002xb,Sommovigo:2003kd}.
}

The two classes of $AdS_3/CFT_2$ pairs are expected to be dual to 1+1-dimensional CFTs whose super-Virasoro algebras 
are, respectively, the small and large ${\cal N}=(4,4)$ superconformal algebras~\cite{Seiberg:1999xz,Gauntlett:1998kc,de Boer:1999rh}. These infinite-dimensional symmetry 
algebras have finite-dimensional Lie sub-superalgebras $psu(1,1|2)^2$ and $d(2,1;\alpha)^2$, where $\alpha=\cos^2\phi$. It is expected that the $CFT_2$ dual of $AdS_3\times S^3\times T^4$ is a deformation of the $Sym^N(T^4)$ sigma-model~\cite{Seiberg:1999xz}. Beyond representation-theoretic statements, very little is known about the $CFT_2$ dual of the $AdS_3\times S^3\times S^3\times S^1$ string theory \cite{Gukov:2004ym}. 

In the last few years, integrability has been used to investigate these dual pairs.\footnote{From the string theory point of view the two $AdS_3$ backgrounds could be treated in parallel, and, what is more, the $\alpha\rightarrow 0$ limit of the 
$AdS_3\times S^3\times S^3\times S^1$ theory gives the (partially decompactified) $AdS_3\times S^3\times T^4$ theory.}
It was observed in \cite{Babichenko:2009dk}, that upon picking a suitable $\kappa$-gauge, Type IIB string theory equations of motion on these backgrounds admit a Lax representation and so the theory is classically integrable. The Lax operator was used \cite{Babichenko:2009dk} to write down integral equations known as the finite-gap equations for this system. The finite-gap equations were discretised and an all-loop Bethe ansatz was proposed for the system in \cite{Babichenko:2009dk,OhlssonSax:2011ms}. An integrable spin-chain whose spectrum was described by the weak coupling limit of this all-loop Bethe ansatz was constructed in \cite{OhlssonSax:2011ms,Sax:2012jv}. The all-loop Bethe ansatz has also been obtained from a different direction, by deriving the S-matrix from the symmetries of the theory and writing down the Bethe-ansatz for the associated spin-chain~\cite{Borsato:2012ud,Borsato:2012ss,Borsato:2013qpa}. The near-BMN limit of string theory on $AdS_3$ has been investigated in~\cite{Rughoonauth:2012qd}. One-loop energy corrections have been computed for giant magnons in~\cite{David:2008yk,Abbott:2012dd,Abbott:2013ixa} and for spinning strings in~\cite{Beccaria:2012kb,Beccaria:2012pm}. Worldsheet scattering amplitudes have been calculated in~\cite{Sundin:2012gc,Sundin:2013ypa,Sundin:2013uca} and compared to the S-matrices in~\cite{Borsato:2012ud} as well as in~\cite{David:2010yg,Ahn:2012hw}. The S-matrix crossing relations have been solved in~\cite{Borsato:2013hoa} and compared to the one-loop string computations of~\cite{Beccaria:2012kb,Abbott:2013ixa,Sundin:2013ypa}. Further, unitary methods have been used in~\cite{Bianchi:2013nra,Engelund:2013fja} to study the S-matrix. Integrability has also been investigated in the context of BTZ black-holes~\cite{David:2011iy,David:2012aq}.

It was already observed in \cite{Babichenko:2009dk} that the finite-gap equations (and hence the all-loop Bethe ansatz) captured the dynamics of massive modes, but not the massless modes.\footnote{Because of the presence of integrability, it is expected that the integrable description of massive modes will get modified in a controlled fashion by adding the massless modes, rather than changing the all-loop Bethe Ansatz completely.} In this paper we show how to incorporate these missing massless modes into the finite-gap equations. We begin in section 2 with a brief review of the BMN limit of $AdS_3\times S^3\times S^3\times S^1$. Then, in section 3 we re-examine the way that the Virasoro constraints are imposed on the finite-gap equations. We find that the way the constraints had been imposed previously in the literature (for example in~\cite{Babichenko:2009dk}) is, in general, too strict. We identify the precise condition placed on the finite-gap equations by the Virasoro constraints. We shall refer to this condition as the generalised residue condition (GRC). 
The GRC is generically less restrictive than the condition used in much of the previous literature.\footnote{The $O(4)$ sigma model, which shares some of the features of the $AdS_3$ backgrounds we consider here was investigated in~\cite{Gromov:2006dh}.}

To illustrate the role of the GRC, in section 4 we focus on the bosonic mode of the $AdS_3\times S^3\times S^3\times S^1$ theory {\em not} associated with the $S^1$ direction. We show that classical string solutions that excite this mode satisfy finite-gap equations when the GRC is imposed. On the other hand, these solutions do not satisfy the constraints previously used in the literature, further explaining the absence of massless modes from the old finite-gap equations. Then, in sections 5 and 6 we show how the complete spectrum of string theory on $AdS_3\times S^3\times S^3\times S^1$ in the BMN limit can be reproduced from the finite-gap equations and the GRC condition. We also show that the complete spectrum for string theory on $AdS_3\times S^3\times T^4$ in the BMN limit can also be obtained using the GRC.

In appendices B and C, we show that for the finite-gap equations of the $AdS_5\times S^5$ and $AdS_4\times\mathbb{CP}^3$ backgrounds the GRC reduces to the old conditions imposed previously in the literature. This is to be expected, as it is well known that for those backgrounds the finite-gap equations previously used in the literature do reproduce the complete spectrum. It is only for backgrounds such as the $AdS_3$ cases we investigate here that the GRC does {\em not} reduce to the conditions used in the previous literature.

\section{BMN limit of $AdS_3\times S^3\times S^3\times S^1$}
In this section we will briefly review the BMN limit~\cite{Blau:2002dy,Berenstein:2002jq}  of string theory on $AdS_3\times S^3\times S^3\times S^1$~\cite{Lu:2002kw,Sommovigo:2003kd} and see how the modes of different masses appear.~\footnote{The BMN limit of string theory on $AdS_3\times S^3\times T^4$ is discussed in~\cite{Hikida:2002in,Gomis:2002qi,Gava:2002xb}.} Starting from the metric
\begin{align}
ds^2=R^2\bigg[d\rho^2&-\cosh^2\rho dt^2+\sinh^2\rho d\gamma^2+\frac{1}{\cos^2\phi}\left(d\theta_1^2+\cos^2\theta_1d\psi_1^2+\sin^2\theta_1d\varphi_1^2\right)\nonumber\\&+\frac{1}{\sin^2\phi}\left(d\theta_2+\cos^2\theta_2d\psi_2^2+\sin^2\theta_2d\varphi_2^2\right)+d\chi^2\bigg] \label{fullmetric}\ ,
\end{align}
we change coordinates as follows (with $\zeta$ being any real constant for now):
\begin{align}
&t=x^++\frac{x^-}{R^2}\ , \quad \rho=\frac{\tilde{x}_2}{R}\ , \quad \theta_1=\cos\phi\frac{\tilde{x}_4}{R}\ , \quad \theta_2=\sin\phi\frac{\tilde{x}_6}{R}\ , \quad \chi=\frac{x_8}{R}\ ,\nonumber\\
&\psi_1=\cos\zeta\cos\phi\left(x^+-\frac{x^-}{R^2}\right)-\sin\zeta\cos\phi\frac{x_1}{R}\ , \quad \psi_2=\sin\zeta\sin\phi\left(x^+-\frac{x^-}{R^2}\right)+\cos\zeta\sin\phi\frac{x_1}{R} \label{coordchange}
\end{align}
and keep only the leading term in the limit $R\to\infty$. The metric reduces to
\begin{equation}
ds^2=-4dx^+dx^-+\sum_{i=1}^8m_i^2x_i^2 (dx^+)^2+\sum_{i=1}^8 dx_i^2\ , \label{BMN}
\end{equation}
with
\begin{equation}
(x_2,x_3)=(\tilde{x}_2\cos\gamma,\tilde{x}_2\sin\gamma), \quad (x_4,x_5)=(\tilde{x}_4\cos\varphi_1,\tilde{x}_4\sin\varphi_1), \quad (x_6,x_7)=(\tilde{x}_6\cos\varphi_2,\tilde{x}_6\sin\varphi_2)
\end{equation}
and masses $m_i$, given by
\begin{equation}
m_2=m_3=1, \quad m_4=m_5=\cos\zeta\cos\phi, \quad m_6=m_7=\sin\zeta\sin\phi, \quad m_1=m_8=0\ .
\end{equation}

The parameter $\zeta$ defines a 1-parameter family of metrics obtained from $AdS_3\times S^3\times S^3\times S^1$ via Penrose limits. This freedom comes from the choice of a relative angle between the geodesics in the two $S^3$ factors. Type II string theory on $AdS_3\times S^3\times S^3\times S^1$ preserves 16 supersymmetries. These remain symmetries of the plane wave limit metric $\eqref{BMN}$; in addition for special values of $\zeta$ there are extra supersymmetries \cite{Gauntlett:1998kc}. If we choose $\zeta=\phi$, string theory on $\eqref{BMN}$ preserves 20 supersymmetries~\cite{Lu:2002kw,Sommovigo:2003kd}. From now on, it will be assumed that we are making this choice, and that the BMN limit has masses
\begin{equation}
m_2=m_3=1\ ,\quad m_4=m_5=\cos^2\phi\ ,\quad m_6=m_7=\sin^2\phi\ ,\quad m_1=m_8=0.
\end{equation}

To find the bosonic spectrum of string theory, we  impose conformal gauge $g_{ab}=\eta_{ab}$ and lightcone gauge $x^+=\kappa\tau$. The equation of motion for $x_i$ then becomes
\begin{equation}
(-\partial_\tau^2+\partial_\sigma^2)x_i=\kappa^2m_i^2x_i \label{eombmn}
\end{equation}
and $x^-$ is determined uniquely from the Virasoro constraints, which in this gauge are
\begin{equation}
\partial_\tau x^-=\frac{1}{4\kappa}\sum_i((\partial_\tau x_i)^2+(\partial_\sigma x_i)^2-\kappa^2m_i^2x_i^2), \quad \partial_\sigma x^-=\frac{1}{2\kappa}\sum_i(\partial_\tau x_i)(\partial_\sigma x_i)\ .\label{x^-deriv}
\end{equation}
In lightcone gauge $x^+$ and $x^-$ become non-dynamical variables and the gauge-fixed Hamiltonian is
\begin{equation}
H=\frac{1}{4\pi\alpha'}\int_0^{2\pi}\mathrm{d}\sigma\sum_{i=1}^8\left[(2\pi\alpha')^2p_i^2+(\partial_\sigma x_i)^2+\kappa^2m_i^2x_i^2\right]\ . \label{lcHam}
\end{equation}
Solving the equations of motion $\eqref{eombmn}$, the $x^i$ have the following mode expansion:
\begin{equation}
x^i=X_0^i+\sqrt{\frac{\alpha'}{2}}\sum_{n=1}^{\infty}\frac{1}{\sqrt{\omega_n^i}}\left(a_n^ie^{-i(\omega_n^i\tau+n\sigma)}+a_n^i{}^\dagger e^{i(\omega_n^i\tau+n\sigma)}+\tilde{a}_n^ie^{-i(\omega_n^i\tau-n\sigma)}+\tilde{a}_n^i{}^\dagger e^{i(\omega_n^i\tau-n\sigma)}\right)\ , \label{mode}
\end{equation}
where
\begin{equation}
\omega_n^i=\sqrt{n^2+\kappa^2 m_i^2}\ ,
\end{equation}
and
\begin{equation}
X_0^i=x_0^i\cos\kappa m\tau+\frac{\alpha'}{\kappa m}p_0^i\sin\kappa m\tau
\end{equation}
for massive modes and
\begin{equation}
X_0^i=x_0^i+\alpha'p_0^i\tau+w^i\sigma
\end{equation}
in the massless case $m_i=0$.\footnote{The winding $w$ in the massless mode is only present if the direction associated to the massless mode in the metric is compact.}

We can insert this mode expansion into the lightcone Hamiltonian $\eqref{lcHam}$. Define the zero modes, for the massive case, as
\begin{equation}
a_0^i=\tilde{a}_0^i=\frac{1}{2}\sqrt{\frac{\alpha'}{\kappa m_i}}p_0^i+\frac{i}{2}\sqrt{\frac{\kappa m_i}{\alpha'}}x_0^i\ ,
\end{equation}
then we have
\begin{equation}
H=\sum_{i=1}^8\sum_{n=0}^\infty\omega_n^i N_n^i+\frac{1}{2\alpha'}\left[(\alpha'p_0^1)^2+(w^1)^2+(\alpha'p_0^8)^2+(w^8)^2\right]\ , \label{Hmod}
\end{equation}
with $N_n^i$ the number operator defined as
\begin{equation}
N_n^i=a_n^i{}^\dagger a_n^i+\tilde{a}_n^i{}^\dagger\tilde{a}_n^i\ .
\end{equation}

Now we consider conserved Noether charges. From the independence of the metric on the coordinates $x^+$ and $x^-$ we get conserved charges $P_+$ and $P_-$ upon integrating the conjugate momenta $p_+$ and $p_-$. These are related to more natural charges: the energy $E=i\partial_\tau$, and an angular momentum $J=-i\partial_\eta$ coming from the spatial coordinate
\begin{equation}
\eta=x^+-\frac{x^-}{R^2}\ . \label{eta}
\end{equation}
Then we have
\begin{equation}
P_+=i\partial_+=i(\partial_t+\partial_\eta)=E-J, \quad P_-=i\partial_-=\frac{i}{R^2}(\partial_t-\partial_\eta)=\frac{E+J}{R^2}
\end{equation}
and
\begin{equation}
P_+=\frac{H}{\kappa}=E-J=\frac{1}{\kappa}\sum_{i=1}^8\sum_{n=0}^\infty\omega_n^i N_n^i+\frac{1}{2\alpha'\kappa}\left[(\alpha'p_0^1)^2+(w^1)^2+(\alpha'p_0^8)^2+(w^8)^2\right]\ . \label{yu}
\end{equation}
Since
\begin{equation}
P_-=\int_0^{2\pi}\mathrm{d}\sigma p_-=\frac{1}{\pi\alpha'}\int_0^{2\pi}\mathrm{d}\sigma \partial_\tau x^+=\frac{2\kappa}{\alpha'}\ ,
\end{equation}
we find $E+J=2\sqrt{\lambda}\kappa$, with $\sqrt{\lambda}=\frac{R^2}{\alpha'}$. To leading order in a large $J$ expansion, $E+J\approx 2J$. So writing the right-hand side of $\eqref{yu}$ in terms of $J$ instead of $\kappa$, to leading order we have $\kappa=\frac{J}{\sqrt{\lambda}}$ and so
\begin{equation}
E-J=\sum_{i=1}^8\sum_{n=0}^\infty\sqrt{m_i^2+\frac{\lambda n^2}{J^2}}\ N_n^i+\frac{\sqrt{\lambda}}{2\alpha'J}\left[(\alpha'p_0^1)^2+(w^1)^2+(\alpha'p_0^8)^2+(w^8)^2\right]\ . \label{XX}
\end{equation}

\section{Coset model, quasimomenta and finite-gap equations}

In this section we will review classical integrability of strings on symmetric space cosets and finite-fap equations~\cite{Kazakov:2004qf,Zarembo:2004hp,Beisert:2005bm}.\footnote{For a more complete discussion and further references see the review~\cite{SchaferNameki:2010jy}.} In section 3.1 we write down a Lax connection~\cite{Bena:2003wd} and from this introduce the complex functions called the quasimomenta which encode the dynamics of the system in their analyticity properties. The quasimomenta satisfy so-called finite-gap equations along their branch cuts. In addition, the quasimomenta always have two simple poles. In section 3.2 we examine the residues at these poles using the auxiliary linear problem, and show that the Virasoro constraints appear in the context of the quasimomenta as a condition on these residues. We emphasise that the condition on the residues which is strictly equivalent to the Virasoro constraints is a more general one than the condition which has been assumed to hold in the literature. We will show in the following sections that these new residue conditions are needed to encode the massless modes into the finite-gap equations of string theory on $AdS_3\times S^3\times S^3$.

\subsection{Integrability on symmetric space cosets}

Consider a coset $G/H_0$, where $G$ is a supergroup and $H_0$ a bosonic sub-group, corresponding to a so-called semi-symmetric space~\cite{Serganova:1983vp}. By definition, such spaces have a $\mathbb{Z}_4$ automorphism acting on them, with the automorphism acting as identity on $H_0$. String theory on such cosets is known to be integrable~\cite{Bena:2003wd}. In the case of $AdS_3$ backgrounds we have $G=H\times H$ corresponding to  left- and right-moving sectors of the dual $CFT_2$. For simplicity let us restrict our attention for now to the bosonic sector of the action, where the $\mathbb{Z}_4$ automorphism reduces to a $\mathbb{Z}_2$ automorphism. For bosonic strings in $AdS_3\times S^3\times S^3$ we have $H_0=SU(1,1)\times SU(2)\times SU(2)$. In the general overview in this subsection we mainly follow $\cite{Zarembo:2010yz}$, and refer the reader to references therein.

We consider an element $g\in G$, and the associated Maurer-Cartan one-form in the Lie algebra of $G$, \begin{equation}
j=g^{-1}dg\in\mathfrak{g}\ . 
\end{equation}
Since $G/H$ is a symmetric space, there exists a $\mathbb{Z}_2$ automorphism $\Omega$ acting on $\mathfrak{g}$, under which we can decompose $j$ as $j=j^{(0)}+j^{(2)}$ where $j^{(0)}$ and $j^{(2)}$ belong to, respectively, the $+1$ and $-1$ eigenspaces of $\Omega$. Explicitly we have
\begin{equation}
j^{(0)}=\frac{1}{2}(j+\Omega(j))\ , \quad j^{(2)}=\frac{1}{2}(j-\Omega(j))\ .
\end{equation}
The action is
\begin{equation}
S=\frac{1}{4\pi\alpha'}\int\mathrm{d}^2\sigma\eta^{\alpha\beta}\mathrm{tr}(j_\alpha^{(2)}j_\beta^{(2)})\ , \label{cosetaction}
\end{equation}
where we have already fixed conformal gauge $g^{\alpha\beta}=\eta^{\alpha\beta}$ in the worldsheet metric. The equation of motion for $j^{(2)}$ is
\begin{equation}
\eta^{\alpha\beta}(\partial_\alpha j_\beta^{(2)}+[j_\alpha^{(0)},j_\beta^{(2)}])=0\ , \label{jeom}
\end{equation}
the Maurer-Cartan relation (Bianchi identity) is
\begin{equation}
\partial_\alpha j_\beta-\partial_\beta j_\alpha +\left[j_\alpha,j_\beta\right]=0\ , \label{mc}
\end{equation}
and the Virasoro constraints are
\begin{equation}
\mathrm{tr}\bigg[(j_\tau^{(2)})^2+(j_\sigma^{(2)})^2\bigg]=\mathrm{tr}\bigg[j_\tau^{(2)}j_\sigma^{(2)}\bigg]=0\ . \label{vir}
\end{equation}
We introduce a Lax connection:
\begin{equation}
L_\alpha=j_\alpha^{(0)}+\frac{z^2+1}{z^2-1}j_\alpha^{(2)}-\frac{2z}{z^2-1}\eta_{\alpha\beta}\epsilon^{\beta\gamma}j_\gamma^{(2)}\ , \label{laxdef}
\end{equation}
where $\epsilon^{\alpha\beta}$ is the two-dimensional antisymmetric tensor with $\epsilon^{01}=1$, and the spectral parameter $z$ is an auxiliary complex parameter giving us a family of connections. The equation of motion $\eqref{jeom}$ and the Maurer-Cartan relation $\eqref{mc}$ are equivalent to the flatness of the Lax connection:
\begin{equation}
\partial_{\lbrack\alpha} L_{\beta\rbrack} + L_{\lbrack\alpha}L_{\beta\rbrack}=0\ .
\end{equation}
We define the monodromy matrix as the path ordered exponential of the Lax connection,
\begin{equation}
M(z)=\mathrm{Pexp}\int_0^{2\pi}\mathrm{d}\sigma L_\sigma(z)\ .
\label{monodef}
\end{equation}
The flatness condition on the Lax connection means that we could equivalently define $M(z)$ to be the integral around any closed curve, but it will be simplest in practice to use a curve of constant $\tau$.

Since $L(z)\in\mathfrak{g}$, $M(z)\in G$. If ${H_l}$ is the Cartan basis of $\mathfrak{g}$, then we can diagonalize $M(z)$ by introducing functions $p_l(z)$ such that 
\begin{equation}
M(z)=\mathrm{exp}\left(\sum_{l=1}^Rp_l(z)H_l\right) \label{mdiag}
\end{equation}
in a diagonal basis, where $R$ is the rank of the algebra $\mathfrak{g}$. The functions $p_l(z)$ are called the {\it quasimomenta}. The dynamics of the sigma model $\eqref{cosetaction}$ are encoded in the analyticity properties of the quasimomenta.

The Lax connection has simple poles at $z=\pm1$ but is otherwise analytic. The quasimomenta inherit these poles from the Lax connection, but may also contain branch cuts. For each quasimomentum $p_l$ we introduce a new index $i$ to count the cuts and denote the collection of branch cuts for $p_l$ by $C_{l,i}$. On these cuts we consider the monodromies of the quasimomenta, coming from the way in which the Riemann surfaces of the quasimomenta are collectively joined and the fact that the quasimomenta are only defined up to multiples of $2\pi i n$. The monodromy relations are\footnote{For ordinary square root branch cuts the right-hand side of $\eqref{monodromy}$ would be zero. Without the Cartan matrix, the non-zero right-hand side of $\eqref{monodromy}$ could be understood by the ambiguity of an overall phase in $p_l$. The presence of the Cartan matrix arises from the fact that the monodromy matrix itself is gauge-dependent, and as a consequence of this the quasimomenta are also only defined up to transformations from the Weyl group. See \cite{Babichenko:2009dk} for more details.}
\begin{equation}
A_{lm}\slashed{p}_m(z)=2\pi in_{l,i}, \quad z\in C_{l,i},\quad n_{l,i}\in\mathbb{Z}\ ,\label{monodromy}
\end{equation}
where $A_{lm}$ is the Cartan matrix of the group and 
\begin{equation}
\slashed{p}_l(z)= \lim_{\epsilon\to 0}(p_l(z+\epsilon)+p_l(z-\epsilon)), \quad z\in C_{l,i}\ ,
\end{equation}
with $\epsilon$ a complex number normal to the branch cut.

We can choose to parametrize the residues at the poles by their sum and difference, defining constants $\kappa_l$ and $m_l$ so that as $z\to\pm 1:$
\begin{equation}
p_l=\frac{1}{2}\frac{\kappa_l z+2\pi m_l}{z\mp 1}+\dots
\end{equation}

The quasimomenta posses an inversion symmetry inherited from the action of the automorphism $\Omega$ on the Lax connection. Since $j^{(0)}$ and $j^{(2)}$ are defined by the action of $\Omega$, we get from the definition of the Lax $\eqref{laxdef}$ that
\begin{equation}
\Omega(L_\alpha(z))=L_\alpha\left(\frac{1}{z}\right)\ .
\end{equation}
This uplifts to an inversion on the monodromy matrix
\begin{equation}
\Omega(M(z))=M\left(\frac{1}{z}\right)\ .
\end{equation}
From this we get an inversion symmetry on the quasimomenta determined by the action of the automorphism on the Cartan basis. If we introduce a matrix $S_{lm}$ such that
\begin{equation}
\Omega(H_l)=\sum_{m=1}^R S_{lm}H_m
\end{equation}
then
\begin{equation}
p_l\left(\frac{1}{z}\right)=\sum_{m=1}^R S_{lm}p_m(z)\ . \label{inversion}
\end{equation}

The Noether charges can be found from the quasimomenta by considering either the limit $z\to 0$ or $z\to\infty$ (these limits are related by the inversion symmetry). For $z\to 0$ for example, the Lax connection can be exanded as
\begin{equation}
L_\sigma=j_\sigma^{(0)}-j_\sigma^{(2)}-2zj_\tau^{(2)}+\mathcal{O}(z^2)\ ,\label{laxz0}
\end{equation}
and $j_\tau^{(2)}$, upon integration over $\sigma$, contains the Noether charges. Recall that the equations of motion $\eqref{jeom}$ imply the conserved current equation
\begin{equation}
\partial_\alpha(g\eta^{\alpha\beta}j_\beta^{(2)}g^{-1})=0\ .\label{noetherj}
\end{equation}

As mentioned above, the quasimomenta will generally contain branch cuts. We can obtain a so-called {\it spectral representation} of the quasimomenta in terms of integrals along these branch cuts. We introduce a density function
\begin{equation}
\rho_l(z)=\lim_{\epsilon\to 0}\left(p_l(z+\epsilon)-p_l(z-\epsilon)\right), \quad z\in C_{l,i}\ .
\end{equation}
Then we have the spectral representation of $p_l$:\footnote{This result comes from applying the Cauchy integral formula on an infinite domain to the function  obtained by subtracting the poles from $p_l$, which is analytic outside this contour surrounding all the cuts. $\eqref{spectralrep}$ then follows by shrinking the contour down onto the cuts. In the case that $p_l$ is meromorphic, this argument is clearly no longer valid. But in that case $\eqref{spectralrep}$ still holds with $\rho_l=0$, since in this case subtracting the poles from the quasimomentum gives an entire function, and the only entire function satisfying the inversion symmetry is a constant.}
\begin{equation}
p_l(z)=\frac{\kappa_l z+2\pi m_l}{z^2-1}+p_l(\infty)+\int_{C_{l,i}}\mathrm{d}w\frac{\rho_l(w)}{z-w}\ .\label{spectralrep}
\end{equation}
The spectral representation is derived assuming nothing about $p_l$ except the nature of its poles and branch cuts. However, we also know that the quasimomenta must satisfy the inversion symmetry $\eqref{inversion}$. This places restrictions on $\kappa_l$, $m_l$ and $p_l(\infty)$:
\begin{equation}
S_{lm}\kappa_m=-\kappa_l, \quad S_{lm}m_m=-m_l, \quad S_{lm}p_m(\infty)=p_l(\infty)-2\pi m_l\ .\label{symres}
\end{equation}
For our purposes we will be able to choose the automorphism $\Omega$ such that $S_{lm}=-\delta_{lm}$.\footnote{If we suppress the distinction between the left-moving and right-moving quasimomenta, as we will indeed be doing later, then this is the form the inversion symmetry will take for us when considering bosonic quasimomenta on $SU(1,1)\times SU(2)\times SU(2)$. If we explicitly distinguish the left-moving and right-moving parts then the inversion symmetry also interchanges them.} In this case the first two relations above are immediately satisfied, and the third determines the constant $p_l(\infty)$ to be
\begin{equation}
p_l(\infty)=\pi m_l\ .
\end{equation}

For a function defined in terms of a density integral as in $\eqref{spectralrep}$, we can apply the Sochocki-Plemelj formula \cite{Sochocki,Plemelj} to evaluate the integral when we take $z$ to be on the contour of integration. With the monodromy of the quasimomentum given by equation $\eqref{monodromy}$, we get from the Sochocki-Plemelj formula
\begin{equation}
A_{lm}\dashint_{C_{l,i}}\mathrm{d}w\frac{\rho_m(w)}{z-w}=-A_{lm}\frac{\kappa_m z+2\pi m_m}{z^2-1}-\pi A_{lm}m_m+2\pi n_{l,i}, \quad z\in C_{l,i}\ .\label{fg}
\end{equation}

These are the finite-gap equations of the system. In the next subsection we see how the Virasoro constraints place restrictions on $\kappa_l$ and $m_l$.

\subsection{WKB analysis and the Virasoro constraint}

There is an equivalent setting \cite{Lax:1968fm} in which to define the monodromy matrix and quasimomenta from a flat Lax connection. In this section we introduce this setting and show one use for it: considering how the Virasoro constraints appear at the level of the quasimomenta.

In the so-called auxiliary linear problem, the Lax connection, viewed as a matrix-valued function of the spectral parameter, is taken to act on a vector space of functions $\Psi_i(\sigma,\tau,z)$ through the first order differential equation

\begin{equation}
\sum_{j=1}^N(\delta_{ij}\partial_\sigma-(L_\sigma)_{ij})\Psi_j(\sigma)=0\ . \label{a}
\end{equation}
where $L_\sigma$ is a $N\times N$ matrix.
The monodromy matrix may be obtained through the relation 
\begin{equation}
\Psi_i(\sigma+2\pi,z)=\sum_{j=1}^NM_{ij}(z)\Psi_j(\sigma,z)
\end{equation}
and we use a basis where $M(z)$ is diagonal with the quasimomenta $p_l$ on the diagonal,\footnote{We will see why the index $l$ appears here shortly.} as in $\eqref{mdiag}$ 
\begin{equation}
\Psi_i(\sigma+2\pi,z)=e^{ip_l(z)}\Psi_i(\sigma,z)\ , \label{b}
\end{equation}

We know that the quasimomenta have poles at $z=\pm1$. Let us determine the residues of these poles by solving the auxiliary linear problem $\eqref{a}$ in the limit $z\to\pm1.$ We denote $h=z\mp 1$ in this limit, so that $h$ is a small parameter we can expand in, and define 
\begin{equation}
V=-ihL_\sigma=-i\left(j_\tau^{(2)}\pm j_\sigma^{(2)}\right)+\mathcal{O}(h), \quad h=z\mp1\ .
\end{equation}
Since $L$ has simple poles at $z=\pm 1$, $V$ is a regular function of $h$. We make the Wentzel-Kramers-Brillouin (WKB) ansatz
\begin{equation}
\Psi_i(\sigma,z)=\exp\left(i\frac{S_l(\sigma,h)}{h}\right)\xi_i(h)\ , \label{c}
\end{equation}
with $\partial_\sigma\xi=0$, so the defining equation $\eqref{a}$ of the system becomes the eigenvalue equation
\begin{equation}
\sum_{j=1}^NV_{ij}\Psi_j=(\partial_\sigma S_l)\Psi_i\ . \label{d}
\end{equation}
In other words, $\partial_\sigma S_l$ are the eigenvalues of $V$. There are $R$ such independent eigenvalues, where $R$ is the rank of the group, hence we use the index $l$ running from 1 to $R$.

With the ansatz $\eqref{c}$, equation $\eqref{b}$ is solved by
\begin{equation}
p_l(z)=\frac{1}{h}\left(S_l(\sigma+2\pi,h)-S_l(\sigma,h)\right)=\frac{1}{h}\int_0^{2\pi}\mathrm{d}\sigma\partial_\sigma S_l(\sigma,h)\ . \label{e}
\end{equation}
We now have an expression for the quasimomenta in terms of $S_l(\sigma,h)$. Since 
\begin{equation}
\mathrm{tr}(V^2)=-\mathrm{tr}\left(j_\tau^{(2)}\pm j_\sigma^{(2)}\right)^2+\mathcal{O}(h), \quad h=z\mp 1\ ,
\end{equation}
the Virasoro constraints, $\eqref{vir}$, imply that $\mathrm{tr}(V^2)=0$ to leading order in $h$. Equation $\eqref{d}$ tells us that $\partial_\sigma S_l$ are the eigenvalues of $V$, so the Virasoro constraints imply\footnote{We are assuming here that we are dealing with bosonic quasimomenta only, so that the Cartan matrix can be chosen to be the identity matrix. In section 6 we give the generalised residue conditions for quasimomenta belonging to a supercoset where we need to include the Cartan matrix.}
\begin{equation}
\sum_{l=1}^R (\partial_\sigma S_l)^2=0+\mathcal{O}(h)\ . \label{ssum}
\end{equation}
If we define
\begin{equation}
f_l^\pm(\sigma)=\lim_{h\to 0} \partial_\sigma S(h,\sigma), \quad h=z\mp 1\ ,
\end{equation}
then taking the limit $h\to 0$ of equation $\eqref{e}$ gives the residues of the quasimomenta as integrals of the functions $f_l^\pm$:
\begin{equation}
\frac{1}{2}(\kappa_l\pm 2\pi m_l)=\int_0^{2\pi}\mathrm{d}\sigma f_l^\pm(\sigma)\ , \label{c1}
\end{equation}
while equation $\eqref{ssum}$, which came from the Virasoro constraints, can be written in terms of $f_l^\pm$ as
\begin{equation}
\sum_{l=1}^R(f_l^\pm)^2=0\ \ . \label{c2}
\end{equation}
Thus, the condition that the Virasoro constraints place upon the residues of the quasimomenta can be stated as follows: the residues can be written as integrals in the form $\eqref{c1}$, such that the integrands satisfy equation $\eqref{c2}$. To clarify this further: there are obviously many different functions of $\sigma$ which give the same result upon integration from $0$ to $2\pi$, and so many choices of $f_l^\pm$ such that $\eqref{c1}$ holds. The condition placed on the residues by the Virasoro constraints is that for at least one of these choices, equation $\eqref{c2}$ holds.

If we knew the residues, and wanted to write down functions to represent them via $\eqref{c1}$, the most obvious and simple choice would be to choose the constant functions
\begin{equation}
f_l^\pm(\sigma)=\frac{1}{4\pi}(\kappa_l\pm 2\pi m_l)\ .\label{fconstant}
\end{equation}
Although we can always make this choice to satisfy equation $\eqref{c1}$, it is not in general guaranteed that this choice for $f_l^\pm$ will satisfy the condition $\eqref{c2}$. The Virasoro constraints imply only that one of the many possible choices for $f_l^\pm$ in equation $\eqref{c1}$ satisfies equation $\eqref{c2}$, not that all possible choices do, or that one particular simple choice does. When the constant functions given by equation $\eqref{fconstant}$ do satisfy equation $\eqref{c2}$, then the condition on the residues can be written as
\begin{equation}
\sum_{l=1}^R(\kappa_l\pm 2\pi m_l)^2=0\ . \label{nullres}
\end{equation}
In much of the literature (see \cite{Zarembo:2010yz} for example), it is the condition of equation $\eqref{nullres}$ that has been taken to hold. In the next section we consider explicit sigma model solutions for strings on $AdS_3\times S^3\times S^3$ and their associated quasimomenta. For each solution we will discuss whether the residues satisfy $\eqref{nullres}$ or only the more general condition written in equations $\eqref{c1}$ and $\eqref{c2}$. We will see that solutions containing massless modes do not satisfy $\eqref{nullres}$, but do satisfy the generalised conditions $\eqref{c1}$ and $\eqref{c2}$. This will show explicitly that the generalised residue condition must be used in the finite-gap equations in order to capture the dynamics of the massless modes.

\section{Strings on $\mathbb{R}\times S^1\times S^1\subset AdS_3\times S^3\times S^3$}

In this section we consider solutions on the subspace $\mathrm{R}\times S^1\times S^1\subset AdS_3\times S^3\times S^3$, with the metric
\begin{equation}
ds^2=R^2\bigg[-dt^2+\frac{1}{\cos^2\phi}d\psi_1^2+\frac{1}{\sin^2\phi}d\psi_2^2\bigg]\ . \label{rs1s1}
\end{equation}

This subspace contains the coset massless mode of the spectrum in the BMN limit.\footnote{Not the one which appears simply as the dynamics of the isolated $S^1$.} If we choose to consider solutions in lightcone gauge in this space with the Virasoro constraints solved before quantization, then we are looking at precisely the same BMN massless mode quantization that we considered as part of the full space in section 2. We will look first at solutions in lightcone gauge, and then in static gauge ($t=\kappa\tau$), since this latter gauge features prominently in the finite-gap analysis. As we will see, the choice of gauge will not affect the dynamics of the general solution. Indeed we will check very explicitly that we have the same form of expression for $E-J$ for each.

We will see presently that the quasimomenta on this subspace have a very simple analytic structure; they have no branch points or cuts, only simple poles at $z=\pm1$. This makes it straightforward to write down the most general quasimomenta for any solution on this space and will serve as a guide for how to incorporate this massless mode into the finite-gap equations.

\subsection{Coset representatives and quasimomenta}

In this subsection we will give an explicit coset representation for solutions on the $\mathbb{R}\times S^1\times S^1$ subspace, chosen in such a way that the quasimomenta are particularly simple to compute. We show that the quasimomenta have no branch points or cuts, and so can be written completely in terms of the residues. In particular, we will write down the most general quasimomenta for any solution on this subspace in terms of the numbers $\kappa_l$ and $m_l$, and what $\kappa_l$ and $m_l$ are in terms of a particular coordinate solution $t(\sigma,\tau)$, $\psi_1(\sigma,\tau)$ and $\psi_2(\sigma,\tau)$. We show how the generalised residue conditions $\eqref{c1}$ and $\eqref{c2}$ are clearly equivalent to the Virasoro conditions expressed in terms of the coordinates. Lastly we write down an expression for $E-J$ in terms of $\kappa_l$ and $m_l$, which we will use later when we consider particular solutions to show that the correct massless dispersion relation appears from the quasimomenta of those solutions.

In the bosonic case the most natural choice of group representative $g$ is a direct sum $g=g_0\oplus g_1\oplus g_2$ with $g_0\in SU(1,1)\times SU(1,1)$ and $g_i\in (SU(2)_i)^2$, where $SU(2)_1$, respectively $SU(2)_2$, is the group manifold for the sphere of radius $\frac{1}{\cos^2\phi}$, respectively $\frac{1}{\sin^2\phi}$. In particular, we choose the coset representatives as follows:

\begin{equation}
g_1=\frac{1}{\cos\phi}\mathrm{diag}\left(e^{i\frac{\psi_1}{2}},e^{-i\frac{\psi_1}{2}},e^{i\frac{\psi_1}{2}},e^{-i\frac{\psi_1}{2}}\right) ,\quad g_2=\frac{1}{\sin\phi}\mathrm{diag}\left(e^{i\frac{\psi_2}{2}},e^{-i\frac{\psi_2}{2}},e^{i\frac{\psi_2}{2}},e^{-i\frac{\psi_2}{2}}\right)\ , \label{g1g2}
\end{equation}
and
\begin{equation}
g_0=\left(
\begin{array}{cccc}
\cosh{\frac{t}{2}}&\sinh{\frac{t}{2}}&0&0\\\sinh{\frac{t}{2}}&\cosh{\frac{t}{2}}&0&0\\ 0&0&\cosh{\frac{t}{2}}&-\sinh{\frac{t}{2}}\\ 0&0&-\sinh{\frac{t}{2}}&\cosh{\frac{t}{2}}
\end{array}
\right)\ .\label{g0}
\end{equation}
Then the current $j=g^{-1}dg$ is
\begin{equation}
j=\frac{dt}{2}\left(
\begin{array}{cccc}
0&1&0&0\\1&0&0&0\\0&0&0&-1\\0&0&-1&0
\end{array}
\right) 
\oplus\frac{i}{\cos\phi}\frac{d\psi_1}{2}\left(
\begin{array}{cccc}
1&0&0&0\\0&-1&0&0\\0&0&1&0\\0&0&0&-1
\end{array}
\right)
\oplus\frac{i}{\sin\phi}\frac{d\psi_1}{2}\left(
\begin{array}{cccc}
1&0&0&0\\0&-1&0&0\\0&0&1&0\\0&0&0&-1
\end{array}
\right)\ . \label{jmassless}
\end{equation}

The $\mathbb{Z}_2$ automorphism on the space is defined here as $\Omega(j)=Kj^tK$, where 
\begin{equation}
K=\left(\begin{array}{cccc}
0&0&-1&0\\0&0&0&-1\\1&0&0&0\\0&1&0&0\\
\end{array}\right)\ .
\end{equation}
For all $j$'s given here, this acts as $\Omega(j)=-j$, so $j^{(0)}=\frac{1}{2}(j+\Omega(j))=0$, $j^{(2)}=\frac{1}{2}(j-\Omega(j))=j$.

We can check explicitly that the coset action $\eqref{cosetaction}$ gives us the sigma model action on the metric $\eqref{rs1s1}$,
\begin{equation}
\mathrm{tr}((j^{(2)})^2)=-2\left(-dt^2+\frac{1}{\cos^2\phi}d\psi_1^2+\frac{1}{\sin^2\phi}d\psi_2^2\right)\ .
\end{equation}
Since $j^{(0)}=0$, the Lax connection is ({\it cf.} equation $\eqref{laxdef}$)
\begin{equation}
L_\sigma=\frac{1}{z^2-1}\left((z^2+1)j_\sigma+2zj_\tau\right)\ .
\end{equation}

The Lax connection is given by a direct sum of three matrices, each of which takes the form of a constant matrix multiplied by a function.\footnote{Classical solutions studied
in~\cite{Gromov:2007aq} have a similarly simple Lax connection.} In this case, the path-ordered exponential taking us from the Lax connection to the monodromy matrix, given in equation~(\ref{monodef}), reduces to an ordinary matrix exponential of the integrals of the scalar functions. It is then straightforward to read-off the quasimomenta 
\begin{equation}
p_1(z)=-\frac{1}{2\cos\phi}\frac{1}{z^2-1}\left((z^2+1)\int_0^{2\pi}\mathrm{d}\sigma\partial_\sigma\psi_1(\sigma,\tau=0)+2z\int_0^{2\pi}\mathrm{d}\sigma\partial_\tau\psi_1(\sigma,\tau=0)\right)\ , \label{q}
\end{equation}
\begin{equation}
p_2(z)=-\frac{1}{2\sin\phi}\frac{1}{z^2-1}\left((z^2+1)\int_0^{2\pi}\mathrm{d}\sigma\partial_\sigma\psi_2(\sigma,\tau=0)+2z\int_0^{2\pi}\mathrm{d}\sigma\partial_\tau\psi_2(\sigma,\tau=0)\right)\, \label{w}
\end{equation}
and
\begin{equation}
p_0(z)=\frac{i}{2}\frac{1}{z^2-1}\left((z^2+1)\int_0^{2\pi}\mathrm{d}\sigma\partial_\sigma t(\sigma,\tau=0)+2z\int_0^{2\pi}\mathrm{d}\sigma\partial_\tau t(\sigma,\tau=0)\right)\ . \label{v}
\end{equation}

The quasimomenta can be written in the form of the spectral representation $\eqref{spectralrep}$, but with no cuts 
\begin{equation}
p_l(z)=\frac{\kappa_l z+2\pi m_l}{z^2-1}+\pi m_l \label{0cutrep}
\end{equation}
where
\begin{align}
&\kappa_0=i\int_0^{2\pi}\mathrm{d}\sigma\partial_\tau t\ , \quad &&\kappa_1=-\frac{1}{\cos\phi}\int_0^{2\pi}\mathrm{d}\sigma\partial_\tau\psi_1\ , \quad &&&\kappa_2=-\frac{1}{\sin\phi}\int_0^{2\pi}\mathrm{d}\sigma\partial_\tau\psi_2\ ,\nonumber\\
&2\pi m_0=i\int_0^{2\pi}\mathrm{d}\sigma\partial_\sigma t\ , \quad &&2\pi m_1=-\frac{1}{\cos\phi}\int_0^{2\pi}\mathrm{d}\sigma\partial_\sigma\psi_1\ , \quad &&&2\pi m_2=-\frac{1}{\sin\phi}\int_0^{2\pi}\mathrm{d}\sigma\partial_\sigma\psi_2\ . \label{kappasms}
\end{align}
Since $t$ must be periodic in $\sigma$, we have $m_0=0$. We also get conditions for integer winding modes on $\psi_1$ and $\psi_2$, namely $m_1\cos\phi\in\mathbb{Z}$ and $m_2\sin\phi\in\mathbb{Z}$.

We noted earlier that the $\mathcal{O}(z)$ term in the quasimomenta as $z\to 0$ should give us the Noether charges of the solution ({\it cf.} equations $\eqref{laxz0}$ and $\eqref{noetherj}$). For these solutions we have, as $z\to 0$, 
\begin{equation}
p_l(z)=-\pi m_l-\kappa_l z+\dots
\end{equation}
and we see that $\kappa_l$ are related to the Noether charges defined from the sigma model action, the energy $E$ and angular momenta $J_1$ and $J_2$ given by:
\begin{equation}
E=\frac{R^2}{2\pi\alpha'}\int_0^{2\pi}\mathrm{d}\sigma\partial_\tau t\ , \quad J_1=\frac{R^2}{2\pi\alpha'\cos^2\phi}\int_0^{2\pi}\mathrm{d}\sigma\partial_\tau\psi_1\ , \quad J_2=\frac{R^2}{2\pi\alpha'\sin^2\phi}\int_0^{2\pi}\mathrm{d}\sigma\partial_\tau\psi_2\ ,
\end{equation}
so that
\begin{equation}
\kappa_0=i\frac{2\pi\alpha'}{R^2}E\ , \quad \kappa_1=-\frac{2\pi\cos\phi\alpha'}{R^2}J_1\ , \quad \kappa_2=-\frac{2\pi\sin\phi\alpha'}{R^2}J_2\ .
\end{equation}

The coefficients of higher order terms in the Taylor expansion of the quasimomenta around $z=0$ give higher conserved charges. For these simple solutions in flat space we can easily see what these terms are. At $\mathcal{O}(z^n)$, the quasimomentum $p_l$ is either proportional to $\kappa_l$ or $m_l$, depending on whether $n$ is odd or even. 

We can see for these simple solutions how the Virasoro constraints restrict the residues of the quasimomenta, as discussed in section 2.2. Using equation~(\ref{kappasms}), we can read off the functions $f_l$ whose $\sigma$-integrals are related to the $\kappa_l$ through $\eqref{c1}$
\begin{equation}
f_0=\frac{i}{2}(\partial_\tau\pm\partial_\sigma)t\ ,\quad f_1=-\frac{1}{2\cos\phi}(\partial_\tau\pm\partial_\sigma)\psi_1\ ,\quad f_2=-\frac{1}{2\sin\phi}(\partial_\tau\pm\partial_\sigma)\psi_2\ .\label{frs1s1}
\end{equation}
A straightforward check then confirms how, for $\mathbb{R}\times S^1\times S^1$, the generalised residue conditions $\eqref{c1}$ and $\eqref{c2}$ are equivalent to the Virasoro condition expressed on the coordinates,
\begin{equation}
\left[(\partial_\tau\pm\partial_\sigma)t\right]^2=\frac{1}{\cos^2\phi}\left[(\partial_\tau\pm\partial_\sigma)\psi_1\right]^2+\frac{1}{\sin^2\phi}\left[(\partial_\tau\pm\partial_\sigma)\psi_2\right]^2\ .
\end{equation}
We noted at the end of section 2.2 that the GRC reduces to the previously used condition $\eqref{nullres}$ when the functions $f_l(\sigma)$ are constants. For these solutions on $\mathbb{R}\times S^1\times S^1$, we can see this occurs only when $t$, $\psi_1$ and $\psi_2$ are all linear functions of $\tau$ and $\sigma$ ({\it i.e.} when the zero mode and winding mode are excited but all other excitations are absent).

It is useful at this point to write down a general expression for $E-J$ in terms of the $\kappa_l$. Recall that $J$ was defined as the Noether charge associated with the angle $\eta$ given in $\eqref{eta}$, so in the $\mathbb{R}\times S^1\times S^1$ subspace it is given by
\begin{equation}
J=\frac{R^2}{\alpha'}\int_0^{2\pi}\mathrm{d}\sigma\partial_\tau\eta=\cos^2\phi J_1+\sin^2\phi J_2
\end{equation}
and therefore
\begin{equation}
E-J=\frac{\sqrt{\lambda}}{2\pi}(-i\kappa_0+\cos\phi\kappa_1+\sin\phi\kappa_2)\ . \label{e-j rs1s1}
\end{equation}

\subsection{Solutions in lightcone gauge}

In this subsection we consider solutions in lightcone gauge $x^+=\kappa\tau$. In this gauge, it is most natural to write down a solution in the coordinates $(x^+,x^-,x_1)$ and then switch to the coordinates $(t,\psi_1,\psi_2)$. Just as in section 2, a solution is given uniquely by specifying $x_1$, as $x^-$ is determined by the Virasoro constraints $\eqref{x^-deriv}$. We will look first at a simple example, and then consider the most general mode expansion for $x_1$. When we do so, we will see that imposing the condition $\eqref{nullres}$ on the residues of the quasimomenta would remove every excitation of this massless mode.\footnote{With the exception of the zero-mode and winding which we will discuss later.}

\subsubsection{Simple example}

Consider
\begin{equation}
x_1=\sqrt{\frac{2\alpha'}{n}}\left(a\cos n(\sigma+\tau)+\tilde{a}\cos \tilde{n}(\tau-\sigma)\right)\ , \label{lc simple}
\end{equation}
with $a$ a real constant and $n$ an integer. Then the Virasoro constraints determine $x^-$:
\begin{equation}\
x^-=\frac{\alpha'}{2\kappa}\left[na(\tau+\sigma)+\tilde{n}\tilde{a}(\tau-\sigma)-\frac{a}{4}\sin 2n(\tau+\sigma)-\frac{\tilde{a}}{4}\sin 2\tilde{n}(\tau-\sigma)\right]\ .
\end{equation}
In terms of $t$, $\psi_1$ and $\psi_2$ the solution is
\begin{align}
t&=\kappa\tau+\frac{\alpha'}{2\kappa R^2}\left[na(\tau+\sigma)+\tilde{n}\tilde{a}(\tau-\sigma)-\frac{a}{4}\sin 2n(\tau+\sigma)-\frac{\tilde{a}}{4}\sin 2\tilde{n}(\tau-\sigma)\right]\ , \nonumber\\
\psi_1&=\kappa\tau\cos^2\phi-\cos^2\phi\frac{\alpha'}{2\kappa R^2}\left[na(\tau+\sigma)+\tilde{n}\tilde{a}(\tau-\sigma)-\frac{a}{4}\sin 2n(\tau+\sigma)-\frac{\tilde{a}}{4}\sin 2\tilde{n}(\tau-\sigma)\right]\nonumber\\&-\sin\phi\cos\phi\sqrt{\frac{2\alpha'}{n}}\left(a\cos n(\sigma+\tau)+\tilde{a}\cos \tilde{n}(\tau-\sigma)\right)\ ,\nonumber\\
\psi_2&=\kappa\tau\sin^2\phi-\cos^2\phi\frac{\alpha'}{2\kappa R^2}\left[na(\tau+\sigma)+\tilde{n}\tilde{a}(\tau-\sigma)-\frac{a}{4}\sin 2n(\tau+\sigma)-\frac{\tilde{a}}{4}\sin 2\tilde{n}(\tau-\sigma)\right]\nonumber\\&+\sin\phi\cos\phi\sqrt{\frac{2\alpha'}{n}}\left(a\cos n(\sigma+\tau)+\tilde{a}\cos \tilde{n}(\tau-\sigma)\right)\ .
\end{align}
The quasimomenta for this solution are given in the standard form $\eqref{0cutrep}$, with $\kappa_l$ and $m_l$ found by inserting the above expression for $t$, $\psi_1$ and $\psi_2$ into $\eqref{kappasms}$ to get
\begin{align}
&\kappa_0=2\pi i\left(\kappa+\frac{\alpha'(na+\tilde{n}\tilde{a})}{2\kappa R^2}\right)\ , \quad &&2\pi m_0=\frac{\pi i\alpha'(na-\tilde{n}\tilde{a})}{\kappa R^2}\ ,\nonumber\\
&\kappa_1=-2\pi\cos\phi\left(\kappa-\frac{\alpha'(na+\tilde{n}\tilde{a})}{2\kappa R^2}\right)\ , \quad &&2\pi m_1=\frac{\pi\alpha'\cos\phi(na-\tilde{n}\tilde{a})}{\kappa R^2}\ ,\nonumber\\
&\kappa_2=-2\pi\sin\phi\left(\kappa-\frac{\alpha'(na+\tilde{n}\tilde{a})}{2\kappa R^2}\right)\ , \quad &&2\pi m_1=\frac{\pi\alpha'\sin\phi(na-\tilde{n}\tilde{a})}{\kappa R^2}\ .
\end{align}
We can see explicitly that these do not satisfy the condition $\eqref{nullres}$ that has been previously taken to hold for the residues of the quasimomenta, indeed we have
\begin{equation}
\sum_{l=0}^2(\kappa_l+2\pi m_l)^2=-\frac{16\pi^2\alpha'na}{R^2}, \quad \sum_{l=0}^2(\kappa_l-2\pi m_l)^2=-\frac{16\pi^2\alpha'\tilde{n}\tilde{a}}{R^2}\ .
\end{equation}

We note that in order to have $m_0=0$ here (the condition that $t$ is periodic in $\sigma$), we must have $na=\tilde{n}\tilde{a}$ and hence also $m_1=m_2=0$. From $\eqref{e-j rs1s1}$ we have for this solution:
\begin{equation}
E-J=\frac{\sqrt{\lambda}\alpha'(na+\tilde{n}\tilde{a})}{\kappa R^2}=(na+\tilde{n}\tilde{a})\frac{\sqrt{\lambda}}{J}\ . \label{e-j lc}
\end{equation}
This matches up with the expression $\eqref{XX}$ for the full spectrum in the BMN limit if we have just a single massless excitation, so this solution does indeed correspond to a massless mode as we expected, and the dispersion relation as obtained from the quasimomenta is the correct one for a massless mode. This is our first example of a massless mode solution which satisfies the generalised residue conditions $\eqref{c1}$ and $\eqref{c2}$ but not the conditions $\eqref{nullres}$.

\subsubsection{General massless mode in lightcone gauge}

Now we consider the most general mode expansion for the massless mode $x_1$, as in $\eqref{mode}$.\footnote{We would like to thank Kostya Zarembo for suggesting we consider the most general massless mode.} We take
\begin{align}
x_1&=x_0+\alpha'p_0\tau+w\sigma\nonumber\\&+\sqrt{\frac{\alpha'}{2}}\sum_{n=1}^\infty\frac{1}{\sqrt{n}}\left(a_ne^{-in(\tau+\sigma)}+a_n^\dagger e^{in(\tau+\sigma)}+\tilde{a}_ne^{-in(\tau-\sigma)}+\tilde{a}_n^\dagger e^{in(\tau-\sigma)}\right)\ .
\end{align}
From $x_1$, $x^-$ is determined via the Virasoro constraints, see equation $\eqref{x^-deriv}$. We can then find $t$,$\psi_1$ and $\psi_2$ from $x_1$ and $x^-$ via equation $\eqref{coordchange}$. The expressions are easily obtained but as they are long and we do not need them we will not write them down explicitly. The quasimomenta have the general form given by equation $\eqref{0cutrep}$ so we only need to find $\kappa_l$ and $m_l$, which ({\it cf.} equation $\eqref{kappasms}$) requires only the $\tau$ and $\sigma$ derivatives of $t$, $\psi_1$ and $\psi_2$. These derivatives will have a double sum in the mode expansion\footnote{This follows since the terms in $\eqref{x^-deriv}$ are squares of derivatives of $x_1$.} coming from $x^-$ and a single sum coming from $x_1$. When we integrate over $\sigma$ in $\eqref{kappasms}$ the double sum reduces to a single sum and we pick up only the zero mode contribution from $x_1$. The conclusion is that the quasimomenta for these solutions are given in the simple form $\eqref{0cutrep}$, with $\kappa_l$ and $m_l$ given by
\begin{align}
&\kappa_0=2\pi i\kappa+\frac{i\pi\alpha'}{\kappa R^2}\sum_{n=1}^\infty n(a_na_n^\dagger+\tilde{a}_n\tilde{a}_n^\dagger)+\frac{i\pi(\alpha'^2p_0^2+w^2)}{2\kappa R^2}\nonumber\\
&2\pi m_0=\frac{i\pi\alpha'}{\kappa R^2}\sum_{n=1}^\infty n(a_na_n^\dagger-\tilde{a}_n\tilde{a}_n^\dagger)+\frac{i\pi\alpha'p_0w}{\kappa R^2}\nonumber\\
&\kappa_1=-2\pi\kappa\cos\phi+\frac{\pi\alpha'\cos\phi}{\kappa R^2}\sum_{n=1}^\infty n(a_na_n^\dagger+\tilde{a}_n\tilde{a}_n^\dagger)+\frac{\pi(\alpha'^2p_0^2+w^2)\cos\phi}{2\kappa R^2}+\frac{2\pi\alpha'p_0\sin\phi}{R}\nonumber\\
&2\pi m_1=\frac{\pi\alpha'\cos\phi}{\kappa R^2}\sum_{n=1}^\infty n(a_na_n^\dagger-\tilde{a}_n\tilde{a}_n^\dagger)+\frac{\pi\alpha'p_0w\cos\phi}{\kappa R^2}+\frac{2\pi w\sin\phi}{R}\nonumber\\
&\kappa_2=-2\pi\kappa\sin\phi+\frac{\pi\alpha'\sin\phi}{\kappa R^2}\sum_{n=1}^\infty n(a_na_n^\dagger+\tilde{a}_n\tilde{a}_n^\dagger)+\frac{\pi(\alpha'^2p_0^2+w^2)\sin\phi}{2\kappa R^2}-\frac{2\pi\alpha'p_0\cos\phi}{R}\nonumber\\
&2\pi m_2=\frac{\pi\alpha'\sin\phi}{\kappa R^2}\sum_{n=1}^\infty n(a_na_n^\dagger-\tilde{a}_n\tilde{a}_n^\dagger)+\frac{\pi\alpha'p_0w\sin\phi}{\kappa R^2}-\frac{2\pi w\cos\phi}{R}\ . \label{fullkapm}
\end{align}

We note that the $\sigma$-periodicity of $t$, $m_0=0$, implies the level matching condition
\begin{equation}
\sum_{n=1}^\infty n(a_na_n^\dagger-\tilde{a}_n\tilde{a}_n^\dagger)+p_0w=0\label{tnowind}
\end{equation}
and so
\begin{equation}
m_1=\frac{w\sin\phi}{R}\ , \quad m_2=-\frac{w\cos\phi}{R}\ .\label{mw}
\end{equation}
Hence, the winding modes in $\psi_1$ and $\psi_2$ come from a winding mode in $x_1$, and the conditions $m_1\cos\phi\in\mathbb{Z}$ and $m_2\sin\phi\in\mathbb{Z}$ are both satisfied if 
\begin{equation}
\frac{w\sin\phi\cos\phi}{R}\in\mathbb{Z}\ .
\end{equation}

From $\eqref{e-j rs1s1}$ we get for $E-J$ for this general solution (approximating $\kappa=\frac{J}{\sqrt{\lambda}}$ again)
\begin{equation}
E-J=\frac{\sqrt{\lambda}}{J}\sum_{n=1}^\infty n(a_n^\dagger a_n+\tilde{a}_n^\dagger\tilde{a}_n)+\frac{(\alpha'p_0^2+\frac{w^2}{\alpha'})\sqrt{\lambda}}{2J}+\mathcal{O}\left(\frac{1}{J^2}\right)
\end{equation}
As expected this is precisely the same as the massless part of the BMN expression $\eqref{XX}$.

The above solutions give a clear indication for why we need to generalise the condition on the residues of the quasimomenta from the conventional one given in $\eqref{nullres}$ to the one proposed in $\eqref{c1}$ and $\eqref{c2}$. To see this, we note that for these solutions, the generalised residue condition is explicitly satisfied.\footnote{We saw from the general expressions $\eqref{frs1s1}$ for $f_l^\pm$ for any solution on $\mathbb{R}\times S^1\times S^1$ in our coset parametrisation how equations $\eqref{c1}$ and $\eqref{c2}$ are equivalent to the Virasoro constraints. Hence our solutions satisfies the residue conditions $\eqref{c1}$ and $\eqref{c2}$ by construction. We have also checked explicitly that the functions $f_l^\pm$ for this solution satisfy equation $\eqref{c2}$.} On the other hand, when we compute the sums of squares of residues as in equation $\eqref{nullres}$ we find
\begin{equation}
\sum_{l=0}^2(\kappa_l+2\pi m_l)^2=-\frac{16\pi^2\alpha'}{R^2}\sum_{n=1}^\infty na_n^\dagger a_n \label{z1}
\end{equation}
and
\begin{equation}
\sum_{l=0}^2(\kappa_l-2\pi m_l)^2=-\frac{16\pi^2\alpha'}{R^2}\sum_{n=1}^\infty n\tilde{a}_n^\dagger \tilde{a}_n\ .\label{z2}
\end{equation}
Imposing the conditions $\eqref{nullres}$ would force us to set all of the massless excitations to zero, with the exception of the zero-mode $p_0$ and winding $w$.\footnote{We noted in section 3.2 that the generalised residue conditions $\eqref{c1}$ and $\eqref{c2}$ reduce to the condition $\eqref{nullres}$ precisely when the functions $f_l^\pm$ are constant. In section 4.1 we saw that for our solutions on $\mathbb{R}\times S^1\times S^1$, the functions $f_l^\pm$ are constant whenever the solution is linear in $\tau$ and $\sigma$, see equation $\eqref{frs1s1}$. We will also see this linear solution in static gauge in the next section, but there is one difference between the two gauges. In lightcone gauge, suppose we set $a_n=\tilde{a}_n=0$ for all $n>1$, as is required if the condition $\eqref{nullres}$ holds. Then the condition that $t$ is periodic in $\sigma$, equation $\eqref{tnowind}$, becomes $p_0w=0$. Hence in lightcone gauge, we can have a solution for $x_1$ with the condition $\eqref{nullres}$ holding on the residues of the quasimomenta if we have either only an excited zero-mode, $x_1=\alpha'p_0\tau$, or a winding mode, $x_1=w\sigma$, but not both. In static gauge, $t$ is already periodic in $\sigma$ by the gauge choice, so we don't have this additional restriction.} Ignoring this single exception for now,\footnote{We will return to the subject of why the linear massless modes were also missing in the previous analysis in section 6.} the above equation demonstrates explicitly why in previous finite-gap 
analysis~\cite{Babichenko:2009dk}, the massless mode was not present. On the other hand, the conditions $\eqref{c1}$ and $\eqref{c2}$ are sufficiently general to incorporate all of the massless modes.

\subsection{Solutions in static gauge}

In static gauge, $t=\kappa\tau$, we cannot take the same approach to writing down a general massless mode solution as in the last sub-section. It has been noted 
previously~\cite{Jorjadze:2012iy}, that quantization of string theory in static gauge is in a certain manner half-way between quantization in lightcone gauge and covariant quantization: in $D$ dimensions gauge fixing in static gauge reduces the degrees of freedom to $D-1$, but it is most natural to impose Virasoro after quantization, so there still remains one spurious degree of freedom.

However, for particularly simple solutions in static gauge, it is possible to solve the Virasoro constraints at the classical level fairly simply. If we work in the coordinates $(t,\eta,x_1)$,\footnote{Recall $\eta$ was defined in $\eqref{eta}$.} then we can write down a solution for $x_1$, and write down the Virasoro constraints as
\begin{equation}
(\partial_\tau\pm\partial_\sigma)\eta=\sqrt{((\partial_\tau\pm\partial_\sigma)t)^2-\frac{1}{R^2}((\partial_\tau\pm\partial_\sigma)x_1)^2}=\sqrt{\kappa^2-\frac{1}{R^2}((\partial_\tau\pm\partial_\sigma)x_1)^2}\ . \label{virtemp}
\end{equation}

We can integrate this in principle to find $\eta$, but for a general $x_1$ the resulting $\eta$ will be given as an integral not expressible in terms of standard functions. 

We note that for all solutions in $\mathbb{R}\times S^1\times S^1$ in static gauge, we can immediately give the component $p_0$ of the quasimomentum from $\eqref{v}$ as
\begin{equation}
p_0=\frac{2i\pi\kappa z}{z^2-1}\ ,\label{p0temp}
\end{equation}
which has the general form $\eqref{0cutrep}$ with $\kappa_0=2\pi i\kappa$ and $m_0=0$.

\subsubsection{Linear solution}

Consider first a simple solution linear in $\tau$ and $\sigma$,
\begin{equation}
x_1=\alpha'p_0\tau+w\sigma\ .
\end{equation}
In this case one can solve the Virasoro constraints $\eqref{virtemp}$ explicitly to get
\begin{equation}
\eta=\frac{1}{2}\sqrt{\kappa^2-\frac{(\alpha'p_0+w)^2}{R^2}}(\tau+\sigma)+\frac{1}{2}\sqrt{\kappa^2-\frac{(\alpha'p_0-w)^2}{R^2}}(\tau-\sigma)\ .
\end{equation}
In terms of $\psi_1$ and $\psi_2$ we have
\begin{equation}\psi_1=\cos\phi\bigg[\psi_1^+(\tau+\sigma)+\psi_1^-(\tau-\sigma)\bigg], \quad \psi_2=\sin\phi\bigg[\psi_2^+(\tau+\sigma)+\psi_2^-(\tau-\sigma)\bigg]\ , \label{linear}
\end{equation}
with $\psi_1^\pm$ and $\psi_2^\pm$ constants given by
\begin{align}
&\psi_1^\pm=\frac{1}{2}\cos\phi\left(\sqrt{\kappa^2-\frac{(\alpha'p_0\pm w)^2}{R^2}}\right)-\sin\phi\frac{(\alpha'p_0\pm w)}{R})\nonumber\\
&\psi_2=\frac{1}{2}\sin\phi\left(\sqrt{\kappa^2-\frac{(\alpha'p_0\pm w)^2}{R^2}}\right)+\cos\phi\frac{(\alpha'p_0\pm w)}{R})\ .
\end{align}
The quasimomenta $p_1$ and $p_2$ are again in the form $\eqref{0cutrep}$ with
\begin{equation}
\kappa_i=-2\pi(\psi_i^++\psi_i^-), \quad m_i=-(\psi_i^+-\psi_i^-)
\end{equation}
for $i=1,2$. The condition for integer winding on $\psi_1$ and $\psi_2$ is that $m_1\cos\phi$ and $m_2\sin\phi$ must be integers ({\it cf.} equation $\eqref{kappasms}$).

Inserting this into $\eqref{e-j rs1s1}$ gives
\begin{equation}
E-J=\sqrt{\lambda}\left(\kappa-\frac{1}{2}\sqrt{\kappa^2-\frac{(\alpha'p_0+w)^2}{R^2}}-\frac{1}{2}\sqrt{\kappa^2-\frac{(\alpha'p_0-w)^2}{R^2}}\right)\ .
\end{equation}

Making again the approximation $J=\sqrt{\lambda}\kappa$ to eliminate $J$ and taking only the leading term in a large $J$ expansion gives
\begin{equation}
E-J=\frac{(\alpha'p_0^2+\frac{w^2}{\alpha'})\sqrt{\lambda}}{2J}+\mathcal{O}\left(\frac{1}{J^2}\right)\ ,\label{e-j tempper}
\end{equation}
and we can compare this with $\eqref{e-j lc}$ to see we have the same form for this expression as we did in lightcone gauge.

For this solution, 
\begin{equation}
\sum_{l=0}^2(\kappa_l\pm2\pi m_l)^2=4\pi^2(-\kappa^2+4(\psi_1^\pm)^2+4(\psi_2^\pm)^2)=0\ .
\end{equation}
Recall that in lightcone gauge, the linear terms in the solution also cancelled in the analogous expressions, see equations $\eqref{z1}$ and $\eqref{z2}$. This is in agreement with the observation in section 4.1 that the generalised residue conditions $\eqref{c1}$ and $\eqref{c2}$ reduce to the previously used condition $\eqref{nullres}$ for linear solutions. In section 6 we will say more about these linear massless mode solutions, and why they were not present in the previous analysis of the quasimomenta in the BMN limit. For now we simply remark that the linear solutions are only a small subsector of the full massless spectrum. As we saw in section 4.2.2, all other massless excitations in lightcone gauge are inconsistent with the residue condition $\eqref{nullres}$. In the next subsection we derive the same conclusion for any single periodic solution in static gauge.

\subsubsection{Periodic solution}

Now we consider the same solution for $x_1$ as we looked at in section 4.2, but this time in static gauge,
\begin{equation}
t=\kappa\tau, \quad x_1=\sqrt{\frac{2\alpha'}{n}}\left(a\cos n(\sigma+\tau)+\tilde{a}\cos \tilde{n}(\tau-\sigma)\right)\ .
\end{equation}
$\eta$ is fixed by the Virasoro constraints:
\begin{equation}
(\partial_\tau+\partial_\sigma)\eta=\sqrt{\kappa^2-\frac{8\alpha'na^2}{R^2}\sin^2 n(\tau+\sigma)}\ ,\quad (\partial_\tau-\partial_\sigma)\eta=\sqrt{\kappa^2-\frac{8\alpha'\tilde{n}\tilde{a}^2}{R^2}\sin^2 \tilde{n}(\tau-\sigma)}\ .
\end{equation}
To integrate this we use the following definition of the incomplete elliptic integral of the second kind:\footnote{We use the non-standard notation $\mathbb{E}$ rather than $E$ to avoid confusion with the energy $E$.}
\begin{equation}
\mathbb{E}(\phi,k)=\int_0^\phi\mathrm{d}\theta \sqrt{1-k^2\sin^2\theta}\ ,
\end{equation}
so that
\begin{equation}
\int\mathrm{d}\sigma^+\partial_+\eta =\frac{\kappa}{2n} \mathbb{E}\left(n\sigma^+,\frac{2\sqrt{2\alpha'n}a}{\kappa R}\right)\ ,\quad \int\mathrm{d}\sigma^-\partial_-\eta =\frac{\kappa}{2\tilde{n}} \mathbb{E}\left(\tilde{n}\sigma^-,\frac{2\sqrt{2\alpha'\tilde{n}}\tilde{a}}{\kappa R}\right) \label{g}
\end{equation}
 for $\sigma^\pm=\tau\pm\sigma$, and hence
\begin{equation}
\eta=\frac{\kappa}{2n}\mathbb{E}\left(n(\tau+\sigma),\frac{2\sqrt{2\alpha'n}a}{\kappa R}\right)+\frac{\kappa}{2\tilde{n}}\mathbb{E}\left(\tilde{n}(\tau-\sigma),\frac{2\sqrt{2\alpha'\tilde{n}}\tilde{a}}{\kappa R}\right)\ .
\end{equation}

From $\eta$ and $x_1$ we have $\psi_1$ and $\psi_2$ ({\it cf.} equation $\eqref{coordchange}$), and can take derivatives and then integrate again in order to determine $\kappa_i$ and $m_i$ ({\it cf.} $\eqref{kappasms}$). We get
\begin{eqnarray}
\kappa_1&=&-2\kappa\cos\phi\left[\mathbb{E}\left(\frac{2\sqrt{2\alpha'n}a}{\kappa R}\right)+\mathbb{E}\left(\frac{2\sqrt{2\alpha'\tilde{n}}\tilde{a}}{\kappa R}\right)\right],\nonumber\\ \kappa_2&=&-2\kappa\sin\phi\left[\mathbb{E}\left(\frac{2\sqrt{2\alpha'n}a}{\kappa R}\right)+\mathbb{E}\left(\frac{2\sqrt{2\alpha'\tilde{n}}\tilde{a}}{\kappa R}\right)\right],\nonumber\\
2\pi m_1&=&-2\kappa\cos\phi\left[\mathbb{E}\left(\frac{2\sqrt{2\alpha'n}a}{\kappa R}\right)-\mathbb{E}\left(\frac{2\sqrt{2\alpha'\tilde{n}}\tilde{a}}{\kappa R}\right)\right],\nonumber\\ 
2\pi m_1&=&-2\kappa\sin\phi\left[\mathbb{E}\left(\frac{2\sqrt{2\alpha'n}a}{\kappa R}\right)-\mathbb{E}\left(\frac{2\sqrt{2\alpha'\tilde{n}}\tilde{a}}{\kappa R}\right)\right],
\end{eqnarray}
written using the complete elliptic integral of the second kind
\begin{equation}
\mathbb{E}(k)=\int_0^{\frac{\pi}{2}}\mathrm{d}\theta\sqrt{1-k^2\sin^2\theta}\ .\label{complete E}
\end{equation}

From $\eqref{e-j rs1s1}$ we have
\begin{equation}
E-J=\sqrt{\lambda}\kappa\left[1-\frac{1}{\pi}\mathbb{E}\left(\frac{2\sqrt{2\alpha'n}a}{\kappa R}\right)-\frac{1}{\pi}\mathbb{E}\left(\frac{2\sqrt{2\alpha'\tilde{n}}\tilde{a}}{\kappa R}\right)\right]\ .
\end{equation}

We make again the approximation $J=\sqrt{\lambda}{\kappa}$ and expand to leading order in $J$, using the expansion for the elliptic integral
\begin{equation}
\mathbb{E}(k)=\frac{\pi}{2}-\frac{\pi}{8}k^2+\mathcal{O}(k^4)
\end{equation}
for $k$ small. From this we get
\begin{equation}
E-J=(na^2+\tilde{n}\tilde{a}^2)\frac{\sqrt{\lambda}}{J}+\mathcal{O}\left(\frac{1}{J^2}\right)\ .
\end{equation}
Comparing this to both the lightcone gauge result $\eqref{e-j lc}$ and the previous static gauge result for a linear solution $\eqref{e-j tempper}$ we see again the same form for the expression, confirming that this solutions corresponds to a massless mode in static gauge.

For this solution we have
\begin{align}
&\sum_{l=0}^2(\kappa_l+2\pi m_l)^2=-4\pi^2\kappa^2+16\kappa^2\left[\mathbb{E}\left(\frac{2\sqrt{2\alpha'n}a}{\kappa R}\right)\right]^2\ ,\nonumber\\
&\sum_{l=0}^2(\kappa_l-2\pi m_l)^2=-4\pi^2\kappa^2+16\kappa^2\left[\mathbb{E}\left(\frac{2\sqrt{2\alpha'\tilde{n}}\tilde{a}}{\kappa R}\right)\right]^2\ ,\label{qwe}
\end{align}
and these expressions are not zero unless $na=\tilde{n}\tilde{a}=0$.\footnote{This follows from the fact that the only solutions to $\mathbb{E}(k)=\frac{\pi}{2}$ for real $k$ are $k=\pm1$.} We conclude that these solutions do not satisfy the residue condition $\eqref{nullres}$ and so would not have been part of the conventional finite-gap analysis. They do however satisfy the generalised conditions $\eqref{c1}$ and $\eqref{c2}$ proposed here.\footnote{As before, this is by construction, {\it cf.} equations $\eqref{frs1s1}$ and the discussion in section 4.2.2.}

\section{Massless mode from $SU(1,1)^2\times SU(2)^2\times SU(2)^2$ quasimomenta}

In the previous section we evaluated the quasimomenta for a number of explicit solutions containing massless mode excitations. We saw how the inclusion of the massless mode required quasimomenta whose residues do not satisfy the condition $\eqref{nullres}$, but instead the more general conditions $\eqref{c1}$ and $\eqref{c2}$. In this section, we look at how using this generalised residue condition, one can derive the presence of the massless mode directly from the finite-gap equations. Later, in section 6.2, we will show how the complete massive and massless spectrum in the BMN limit can be derived from the $D(2,1;\alpha)^2$ finite-gap equations. As such we will focus on the massless modes in this section. We will show that using equation $\eqref{e-j rs1s1}$ for $E-J$ in terms of the residues together with the GRC, is is possible to derive the presence of the massless excitation.

In \cite{Babichenko:2009dk} the residues had been chosen to be\footnote{That is, $\eqref{oldres}$ is equivalent to the choice of residues in \cite{Babichenko:2009dk} once one allows for the restriction of $D(2,1;\alpha)$ to its bosonic subgroup and the appropriate changes in grading and gauge choices.}
\begin{equation}
\kappa_0=2\pi i\kappa, \quad \kappa_1=-2\pi\kappa\cos\phi, \quad \kappa_2=-2\pi\kappa\sin\phi\ ,\label{oldres}
\end{equation}
so that
\begin{equation}
-i\kappa_0+\cos\phi\kappa_1+\sin\phi\kappa_2=0\ .
\end{equation}
Here, we do not make this assumption. Instead we require that the residues be given as integrals of functions as in equation $\eqref{c1}$ with the integrands obeying equation $\eqref{c2}$. The only singularities of the BMN vacuum quasimomenta are poles with residues as in equation $\eqref{oldres}$. Hence when we consider solutions in the BMN limit, the residues will be given by equation $\eqref{oldres}$ to leading order in $\kappa$.\footnote{The BMN limit involves taking $J$ large. $\kappa$ is proportional to $J$ to leading order and we will ultimately be interested only in the leading term in the expressions we derive. Hence, we can consider a large $\kappa$ expansion.} This leading term gives no contribution to the expression for $E-J$, so we are interested in finding the highest order term that does contribute. Our approach will thus be to consider a large $\kappa$ expansion for the most general residues which firstly satisfy the condition $\eqref{c1}$ and $\eqref{c2}$, and secondly are given by equation $\eqref{oldres}$ to leading order.

For simplicity we set the winding parameters $m_l$ to zero. Then the functions $f_l^\pm$ in $\eqref{c1}$ obey $f_l^+=f_l^-$ and we denote them by $f_l$, with
\begin{equation}
\kappa_l=\int_0^{2\pi}\mathrm{d}\sigma f_l(\sigma)\ .
\end{equation}
Since we are taking a large $\kappa$ expansion, we will also henceforth put in explicit dependence of $\kappa$ whenever it appears, so $f_l=f_l(\sigma,\kappa)$. We can solve the condition $\eqref{c2}$ on the functions $f_l$ by introducing a new function $\zeta(\sigma,\kappa)$ such that
\begin{equation}
f_1(\sigma,\kappa)=i\cos\zeta(\sigma,\kappa)f_0(\sigma,\kappa), \quad f_2(\sigma,\kappa)=i\sin\zeta(\sigma,\kappa)f_0(\sigma,\kappa)\ .\label{solvedfvir}
\end{equation}

We fix the leading term of  $f_0$ in the large $\kappa$ expansion to give the BMN vacuum value for $\kappa_0$ in equation $\eqref{oldres}$ and leave lower order terms undetermined:
\begin{equation}
f_0(\sigma,\kappa)=i\kappa+if_0^0(\sigma)+i\frac{1}{\kappa}f_0^1(\sigma)+\mathcal{O}\left(\frac{1}{\kappa^2}\right)\ .\label{f0expansion}
\end{equation}
Then, with $f_1$ and $f_2$ given in terms of $\zeta$ and $f_0$ through equation $\eqref{solvedfvir}$, we get the correct leading order terms for $\kappa_1$ and $\kappa_2$ provided $\zeta(\sigma,\kappa)$ is equal to $\phi$ to leading order in $\kappa$. In particular, we expand $\zeta$ with the first term fixed and all subsequent terms arbitrary function of $\sigma$:
\begin{equation}
\zeta(\sigma,\kappa)=\phi+\frac{1}{\kappa}\zeta^1(\sigma)+\frac{1}{\kappa^2}\zeta^2(\sigma)+\mathcal{O}\left(\frac{1}{\kappa^3}\right)\ .\label{zetaexpansion}
\end{equation}
Inserting the expansions for $\zeta$ and $f_0$ into equation $\eqref{solvedfvir}$, we find
\begin{align}
f_1(\sigma,\kappa)&=-\kappa\cos\phi+\sin\phi\ \zeta^1(\sigma)-\cos\phi\ f_0^0(\sigma)\nonumber\\
&+\frac{1}{\kappa}\left[\sin\phi\ \zeta^2(\sigma)+\frac{1}{2}\cos\phi\ \zeta^1(\sigma)^2+\sin\phi\ \zeta^1(\sigma)f_0^0(\sigma)-\cos\phi\ f_0^1(\sigma)\right]+\mathcal{O}\left(\frac{1}{\kappa^2}\right)\ ,\label{f1expansion}
\end{align}
\begin{align}
f_2(\sigma,\kappa)&=-\kappa\sin\phi-\cos\phi\ \zeta^1(\sigma)-\sin\phi\ f_0^0(\sigma)\nonumber\\
&+\frac{1}{\kappa}\left[-\cos\phi\ \zeta^2(\sigma)+\frac{1}{2}\sin\phi\ \zeta^1(\sigma)^2-\cos\phi\ \zeta^1(\sigma)f_0^0(\sigma)-\sin\phi\ f_0^1(\sigma)\right]+\mathcal{O}\left(\frac{1}{\kappa^2}\right)\ .\label{f2expansion}
\end{align} 
When we insert the expansions of $f_l$ given in equations $\eqref{f0expansion}$, $\eqref{f1expansion}$ and $\eqref{f2expansion}$ into equation $\eqref{e-j rs1s1}$ for $E-J$, we find that not only do the terms of $\mathcal{O}(\kappa)$ cancel, as we knew they should (since we fixed the leading order terms to be the BMN vacuum), but also the terms of $\mathcal{O}(1)$ cancel. This is precisely what is required for the extra mode coming from the residues to be massless.~\footnote{To see this, note that the right-hand side of equation $\eqref{XX}$ is $\mathcal{O}(1)$ for massive modes, but $\mathcal{O}\left(\frac{1}{J}\right)$ for the massless mode.} In particular, we find
\begin{equation}
-if_0(\sigma,\kappa)+\cos\phi f_1(\sigma,\kappa)+\sin\phi f_2(\sigma,\kappa)=\frac{1}{2\kappa}\zeta^1(\sigma)^2+\mathcal{O}\left(\frac{1}{\kappa^2}\right)\ .
\end{equation}

The final step in deriving the massless spectrum uses the observation that as the functions $f_l$ are eigenvalues of the Lax connection $L_\sigma$, which is a periodic function of $\sigma$,\footnote{The coset representative $g\in SU(1,1)^2\times SU(2)^2\times SU(2)^2$ should be periodic in $\sigma$ for closed strings.} $f_l$ are also periodic functions of $\sigma$ and hence so is $\zeta^1$. Other than this, $\zeta^1$ is an arbitrary function, so we can write it in a mode expansion (with the normalisations chosen for our convenience):
\begin{equation}
\zeta^1(\sigma)=\sqrt{\alpha'}\lambda^{-\frac{1}{4}}p_0+\sqrt{2}\lambda^{-\frac{1}{4}}\sum_{n=1}^\infty\sqrt{n}\left(a_ne^{-in\sigma}+a_n^\dagger e^{in\sigma}\right)\ .\label{zetamode}
\end{equation}
Then the contribution to $E-J$ from the residues is
\begin{equation}
E-J=\frac{\sqrt{\lambda}}{4\pi\kappa}\int_0^{2\pi}\mathrm{d}\sigma\zeta^1(\sigma)^2+\mathcal{O}\left(\frac{1}{\kappa^2}\right)=\frac{\sqrt{\lambda}}{J}\left(\frac{\alpha' p_0^2}{2}+\sum_{n=1}^\infty na_n^\dagger a_n\right)+\mathcal{O}\left(\frac{1}{J^2}\right)\ ,
\end{equation}
which is the full contribution to the spectrum in the BMN limit from the massless mode $x_1$ in $\eqref{XX}$.\footnote{Apart from the winding mode $w$, which we neglected by setting $m_l=0$ earlier in this section. From $\eqref{mw}$ we can see directly that in lightcone gauge, $m_l=0$ implies $w=0$. Though less obvious, the same statement can be confirmed to be true for the linear solution in static gauge. Including the winding does not alter the analysis in any way, but requires the functions $f_l^+$ and $f_l^-$ to be kept distinct, so we have ignored it here  to keep the notation simpler. Note also that we only defined a mode expansion for $\zeta^1$ in terms of $a_n$ and neglected a corresponding $\tilde{a}_n$, again this is to keep the notation simple, and because the level-matching condition allows us to write $E-J$ solely in terms of contributions from left-movers when $w=0$, see equation $\eqref{tnowind}$.}

Finally, we can return to the question of the linear massless mode seen in section 4 in both lightcone and static gauges, and ask why it was not seen in previous analysis even though its residues do satisfy the previously used residue condition $\eqref{nullres}$. The answer is that the assumptions made in previous work have not been solely to impose the condition $\eqref{nullres}$, but to take the residues to be precisely those of the BMN vacuum, namely as in equation $\eqref{oldres}$. In particular this implies $\zeta^1(\sigma)=0$. This is a stronger condition still than $\partial_\sigma\zeta^1(\sigma)=0$, which is what follows from the residue condition $\eqref{nullres}$. Generalising the residues beyond the BMN vacuum values but keeping the residue condition $\eqref{nullres}$ would add the zero-mode\footnote{It would also add the winding term if we included it.} to $\zeta^1$ and hence a single massless excitation. 

\section{Finite-gap equations and generalised residue conditions}

So far in this paper we have focused our attention on quasimomenta for bosonic strings only.  It is straightforward to find the generalisation of the GRC for finite-gap equations on a supercoset. The  residues of the quasimomenta are still given by equation~\eqref{c1} but now the  functions $f_l^\pm(\sigma)$ satisfy 
\begin{equation}
\sum_{l,m} A_{lm} f_l^\pm f_m^\pm=0\ ,\label{d21acon}
\end{equation}
where $A_{l,m}$ is the Cartan matrix of the supergroup.

 Although the generalised residue condition of equations $\eqref{c1}$ and $\eqref{d21acon}$ is the correct residue condition to use for strings on any supercoset, there are supercosets for which this condition is equivalent to the residue condition used widely in the literature
\begin{equation}
\sum_{l,m} A_{lm}(\kappa_l\pm2\pi m_l)(\kappa_m\pm2\pi m_m)=0\ .\label{cartanoldcon}
\end{equation}
Specifically, we show in appendices B and C that the above residue condition is equivalent to the GRC for strings on $AdS_5\times S^5$ and $AdS_4\times\mathbb{CP}^3$. This was to be expected since for those backgrounds the conventional finite-gap equations are well known to capture the complete string spectrum.

In the rest of this section we will look at the implications of the GRC for quasimomenta on $AdS_3$ backgrounds. First, in section 6.1 we write down the finite-gap equations with generalised residues for superstrings on $AdS_3\times S^3\times S^3\times S^1$. In section 6.2 we show that these finite-gap equations with the GRC reproduce the complete (massive {\em and} massless) BMN spectrum for this background. In section 6.3 we investigate the $AdS_3\times S^3\times T^4$  finite-gap equations with GRC and show that we can similarly incorporate all massless modes into the finite-gap equations for that system.~\footnote{We are grateful to Kostya Zarembo for discussions on the way that the free bosons enter this analysis.}

\subsection{$D(2,1;\alpha)^2\times U(1)^2$ finite-gap equations}

We use a subscript $\pm$ to refer to the left and right sectors of the supergroup, $D(2,1;\alpha)_+\times D(2,1;\alpha)_-$. The Cartan matrix for this supergroup takes the form
\begin{equation}
A=\left(
\begin{array}{cccc}
4\sin^2\phi & -2\sin^2\phi & 0\\
-2\sin^2\phi & 0 & -2\cos^2\phi\\
0 & -2\cos^2\phi & 4\cos^2\phi
\end{array}\right)
\otimes \mathbf{1}_2\ .\label{d21acartan}
\end{equation}
The $D(2,1;\alpha)_+\times D(2,1;\alpha)_-$ quasimomenta are $p_l^\pm$ where $l=1,2,3$.\footnote{It is no longer natural to use the notation $l=0,1,2$ as we did in the bosonic subgroup as the quasimomenta are no longer associated naturally to block diagonal subalgebras with either Lorentzian or Euclidean signature.} The identity factor in $A$ is a $2\times 2$ identity matrix acting on the $\pm$ indices. The action of the inversion symmetry on the quasimomenta is given by equation $\eqref{inversion}$ with
\begin{equation}
S=\mathbf{1}_3\otimes \sigma^1\ .\label{d21as}
\end{equation}

When writing down the spectral representation $\eqref{spectralrep}$ for the quasimomenta on this space, it is convenient to use the inversion symmetry to write the integrals over cuts inside the unit circle in terms of the integrals over cuts outside the unit circle. Once we take account of the necessary effect of the symmetry on the density function $\rho_l(z)$, the spectral representation can then be written as
\begin{equation}
p_l^\pm(z)=\frac{\kappa_l^\pm z+2\pi m_l^\pm}{z^2-1}+\pi m_l^\pm+\int\mathrm{d}w\frac{\rho_l^\pm(w)}{z-w}+\int\frac{\mathrm{d}w}{w^2}\frac{\rho_l^\mp(w)}{z-\frac{1}{w}}\ ,
\end{equation}
where all integrals are over cuts outside the unit circle, and we have given the same index structure to the densities $\rho_l^\pm$ and the residues $\kappa_l^\pm$ and $m_l^\pm$. In fact, $\kappa_l^+$ is simply related to $\kappa_l^-$ by the inversion symmetry (and similarly $m^-$ to $m^+$), see equation $\eqref{symres}$
\begin{equation}
\kappa^+_l=-\kappa^-_l\,,\qquad 
m^+_l=-m^-_l\,.
\end{equation}
The finite-gap equations for $D(2,1;\alpha)_+\times D(2,1;\alpha)_-$ are then given as follows:
\begin{align}
\mp&4\sin^2\phi\frac{\kappa_1z+2\pi m_1}{z^2-1}\pm2\sin^2\phi\frac{\kappa_2z+2\pi m_2}{z^2-1}+2\pi n_{1,i}^\pm\nonumber\\&=4\sin^2\phi\dashint\mathrm{d}w\frac{\rho_1^\pm(w)}{z-w}-2\sin^2\phi\dashint\mathrm{d}w\frac{\rho_2^\pm(w)}{z-w}-4\sin^2\phi\int\frac{\mathrm{d}w}{w^2}\frac{\rho_1^\mp(w)}{z-\frac{1}{w}}+2\sin^2\phi\int\frac{\mathrm{d}w}{w^2}\frac{\rho_2^\mp(w)}{z-\frac{1}{w}}\label{fg1}
\end{align}
\begin{align}
\pm&2\sin^2\phi\frac{\kappa_1z+2\pi m_1}{z^2-1}\pm2\cos^2\phi\frac{\kappa_3z+2\pi m_3}{z^2-1}+2\pi n_{2,i}^\pm\nonumber\\&=-2\sin^2\phi\dashint\mathrm{d}w\frac{\rho_1^\pm(w)}{z-w}-2\cos^2\phi\dashint\mathrm{d}w\frac{\rho_3^\pm(w)}{z-w}+2\sin^2\phi\int\frac{\mathrm{d}w}{w^2}\frac{\rho_1^\mp(w)}{z-\frac{1}{w}}+2\cos^2\phi\int\frac{\mathrm{d}w}{w^2}\frac{\rho_3^\mp(w)}{z-\frac{1}{w}}\label{fg2}
\end{align}
\begin{align}
\pm&2\cos^2\phi\frac{\kappa_2z+2\pi m_2}{z^2-1}\mp4\cos^2\phi\frac{\kappa_3z+2\pi m_3}{z^2-1}+2\pi n_{l,i}^\pm\nonumber\\&=4\cos^2\phi\dashint\mathrm{d}w\frac{\rho_3^\pm(w)}{z-w}-2\cos^2\phi\dashint\mathrm{d}w\frac{\rho_2^\pm(w)}{z-w}-4\cos^2\phi\int\frac{\mathrm{d}w}{w^2}\frac{\rho_3^\mp(w)}{z-\frac{1}{w}}+2\cos^2\phi\int\frac{\mathrm{d}w}{w^2}\frac{\rho_2^\mp(w)}{z-\frac{1}{w}}\ .\label{fg3}
\end{align}

For the $U(1)^2$ part of the theory, the situation is much simpler. We have just the additional quasimomenta  $p_4^\pm$. The Cartan matrix can be taken to be the identity while the inversion matrix is $S=\sigma^1$, {\it i.e.} it  interchanges $p_4^+$ and $p_4^-$. Both the Cartan matrix and inversion matrix for the full theory are direct sums of the $D(2,1;\alpha)^2$ terms given above with the simple $U(1)^2$ terms. Clearly, $p_4^\pm$ trivially satisfy their own finite-gap equations with no cuts.

The residues $\kappa_l\pm 2\pi m_l$ are written in terms of functions $f_l^\pm(\sigma)$ ({\it cf.} equation $\eqref{c1}$),\footnote{Note that the $\pm$ index on $f_l^\pm$ refers to $\kappa\pm 2\pi m_l$ and is not the same as the $\pm$ index on $p_l^\pm$.} and these functions $f_l^\pm$ satisfy equation $\eqref{d21acon}$. With the inversion symmetry satisfied (so that we can write the residues of the right-movers in terms of the left-movers say), the GRC is
\begin{equation}
\sum_{l,m=1}^3 A_{lm}f_l^\pm f_m^\pm + (f_4^\pm)^2=0\ ,
\end{equation}
where $A_{lm}$ here denotes just the $3\times 3$ Cartan matrix in equation $\eqref{d21acartan}$. Explicitly this is
\begin{equation}
4\sin^2\phi\left(f_1^\pm-\frac{1}{2}f_2^\pm\right)^2+4\cos^2\phi\left(f_3^\pm-\frac{1}{2}f_2^\pm\right)^2+(f_4^\pm)^2=(f_2^\pm)^2\ .
\end{equation}
Whereas in section 5 we solved the condition by introducing functions $\zeta^\pm(\sigma)$, now we also introduce a second new pair of functions $\chi^\pm(\sigma)$ and write the solution to this condition as
\begin{align}
&2\sin\phi\left(f_1^\pm-\frac{1}{2}f_2^\pm\right)=-\sin\zeta^\pm\cos\chi^\pm\ f_2^\pm\nonumber\\
&2\cos\phi\left(f_3^\pm-\frac{1}{2}f_2^\pm\right)=-\cos\zeta^\pm\cos\chi^\pm\ f_2^\pm,\nonumber\\&f_4^\pm=\sin\chi^\pm f_2^\pm\ .\label{f1f3-f2zeta}
\end{align}

Therefore, the complete proposal for the finite-gap equations with the generalised residue condition is given by equations $\eqref{fg1}$, $\eqref{fg2}$ and $\eqref{fg3}$, with $\kappa_l$ and $m_l$ given in terms of $f_l^\pm$ via equation $\eqref{c1}$, and $f_1^\pm$,$f_3^\pm$ and $f_4^\pm$ written in terms of $f_2^\pm$ and additional functions $\zeta^\pm$ and $\chi^\pm$ via equation $\eqref{f1f3-f2zeta}$. 

\subsection{Matching the full BMN spectrum of $D(2,1;\alpha)^2\times U(1)^2$}

In this subsection we show how the above finite-gap equations and GRC can be used to derive the  BMN limit of the spectrum of superstrings on $AdS_3\times S^3\times S^3\times S^1$. For simplicity we will neglect the winding $m^\pm_l$, so that $f_l^+=f_l^-$, and we denote $f_l=f_l^+=f_l^-$. Expanding in $z$ we obtain the following expression for $E-J$ 
\begin{equation}
E-J=\frac{\sqrt{\lambda}}{2\pi}\left[2\sin^2\phi\kappa_1+2\cos^2\phi\kappa_3+\sum_{s=\pm}s\left(\sin^2\phi\int_{C_{1,i}}\mathrm{d}w\frac{\rho_1^s(w)}{w^2}+\cos^2\phi\int_{C_{3,i}}\mathrm{d}w\frac{\rho_3^s(w)}{w^3}\right)\right]\ . \label{E-J full}
\end{equation}
Notice that $p_2$ and $p_4$ do not contribute to $E-J$.
For the BMN vacuum the $f_l$ are~\footnote{These are the values which are taken in \cite{Babichenko:2009dk} for all states, not just the BMN vacuum.}
\begin{equation}
f_1=f_3=f_4=0, \quad f_2=\kappa\ .\label{d21aleading}
\end{equation}
Next we make an expansion around the BMN vacuum by expanding in large $\kappa$, with the leading order terms in $f_l$ given by equation $\eqref{d21aleading}$. There is no $\mathcal{O}(\kappa)$ term for $f_1$ and $f_3$, as in equation $\eqref{d21aleading}$, provided that the leading order term in $\zeta$ is $\phi$, just as we had in equation $\eqref{zetaexpansion}$. As pointed out below equation $\eqref{f2expansion}$ this is to be expected of massless modes. There is no $\mathcal{O}(\kappa)$ term for $f_4$ provided that $\chi\to0$ for large $\kappa$ We therefore make exactly the same expansion for $\zeta$ as in equation $\eqref{zetaexpansion}$, and the following expansion for $f_2$ and $\chi$:
\begin{equation}
f_2(\sigma,\kappa)=\kappa+f_2^0(\sigma)+\frac{1}{\kappa}f_2^1(\sigma)+\mathcal{O}\left(\frac{1}{\kappa^2}\right)\ ,\quad \chi(\sigma,\kappa)=\frac{1}{\kappa}\chi^1(\sigma)+\mathcal{O}\left(\frac{1}{\kappa^2}\right)\ .
\end{equation}
Then $f_1$ and $f_3$ have the following expansions:
\begin{align}
f_1(\sigma,\kappa)&=-\frac{1}{2}\cot\phi\ \zeta^1(\sigma)\nonumber\\&+\frac{1}{2\kappa}\left(-\cot\phi\ \zeta^2(\sigma)+\frac{1}{2}\zeta^1(\sigma)^2-\cot\phi\ \zeta^1(\sigma)f_2^0(\sigma)+\frac{1}{2}\chi^1(\sigma)^2\right)+\mathcal{O}\left(\frac{1}{\kappa^2}\right)\ ,\label{f1exp}
\end{align}
\begin{align}
f_3(\sigma,\kappa)&=\frac{1}{2}\tan\phi\ \zeta^1(\sigma)\nonumber\\&+\frac{1}{2\kappa}\left(\tan\phi\ \zeta^2(\sigma)+\frac{1}{2}\zeta^1(\sigma)^2+\tan\phi\ \zeta^1(\sigma)f_2^0(\sigma)+\frac{1}{2}\chi^1(\sigma)^2\right)+\mathcal{O}\left(\frac{1}{\kappa^2}\right)\ \label{f3exp}
\end{align}
and from this we get
\begin{equation}
\sin^2\phi\ f_1(\sigma,\kappa)+\cos^2\phi\ f_3(\sigma,\kappa)=\frac{1}{4\kappa}\left(\zeta^1(\sigma)^2+\chi^1(\sigma)^2\right)+\mathcal{O}\left(\frac{1}{\kappa^2}\right)\ .
\end{equation}
The expansion for $f_4$ meanwhile is
\begin{equation}
f_4(\sigma,\kappa)=\chi^1(\sigma)+\frac{1}{\kappa}\chi^1(\sigma)f_2^0(\sigma)+\mathcal{O}\left(\frac{1}{\kappa^2}\right)\ .
\end{equation}
As in section 5, we can construct a massless boson from $\zeta^1$ in the following way. Since $\zeta^1$ is a periodic function, we make a mode expansion for it as in equation $\eqref{zetamode}$, and inserting this into equation $\eqref{E-J full}$ gives us the spectrum of a single massless boson. We can do exactly the same for $\chi^1$ with a second bosonic mode expansion which gives us a second boson. These two bosons can be distinguished by the fact that $\chi^1$ appears in the expansion for $f_4$ while $\zeta^1$ does not, therefore only one of the bosons is charged under the $U(1)$ associated to translations along $S^1$.

We have seen how the massless bosonic modes now appear in the analysis of the full $D(2,1;\alpha)^2\times U(1)^2$ finite-gap equations. The bosonic modes of mass $\cos^2\phi$ and $\sin^2\phi$ are found by the same procedure as in \cite{Babichenko:2009dk}. We simply have to add one additional step at the start of the procedure: to identify a single massive mode only, we take only the leading, BMN vacuum, term in the expansion for the residues, see equation $\eqref{d21aleading}$. Then we also neglect the integral terms of the right-hand side of the finite-gap equations $\eqref{fg1}$, $\eqref{fg2}$ and $\eqref{fg3}$ in order to take the BMN limit. Taking equation $\eqref{fg1}$ in this way gives the mode of mass $\cos^2\phi$, equation $\eqref{fg3}$ gives the mode of mass $\sin^2\phi$, and equation $\eqref{fg2}$ does not contribute to the massive modes. The mode of mass 1 appears as a stack of the other two massive modes \cite{Beisert:2005di,Gromov:2007ky,Babichenko:2009dk}.

Next we obtain the massless fermions. The situation is closely analogous to that for the massive modes. The bosonic mode of mass $\sin^2\phi$ say, appears in the BMN limit of a solution whose only non-trivial quasimomentum is $p_1$, corresponding to a bosonic link in the Dynkin diagram. The fermion of the same mass then appears as a stack going from $p_1$ to $p_2$, the quasimomentum corresponding to a fermionic link. We have seen how one massless boson appears when we make a mode expansion for the parameter $\zeta^1(\sigma)$ which appears in the expansion around the BMN vacuum of a solution to the generalised residue conditions ({\it cf.} $\eqref{zetaexpansion}$). If this is the only term in the expansions that we make non-zero, except for the leading order, vacuum terms, then we have an excitation which appears in the residues $\kappa_1$ and $\kappa_3$, but not $\kappa_2$. We can produce a fermion by turning on terms which also contribute to $\kappa_2$. In particular we choose a solution with $\zeta^1(\sigma)=f_2^0(\sigma)$ in close analogy with the massive fermions. We then make a fermionic mode expansion similarly to the bosonic mode expansion $\eqref{zetamode}$:
\begin{equation}
\zeta^1(\sigma)=f_2^0(\sigma)=\sqrt{\alpha'}\lambda^{-\frac{1}{4}}\psi_0+\sqrt{2}\lambda^{-\frac{1}{4}}\sum_{n=1}^\infty\sqrt{n}\left(\psi_ne^{-in\sigma}+\psi_n^\dagger e^{in\sigma}\right)\ .\label{fermionmode}
\end{equation}
Then $E-J$ for this solution is given by
\begin{equation}
E-J=\frac{\sqrt{\lambda}}{J}\left(\frac{\alpha' \psi_0^2}{2}+\sum_{n=1}^\infty n\psi_n^\dagger \psi_n\right)+\mathcal{O}\left(\frac{1}{J^2}\right)\ .
\end{equation}
In other words it contributes to $E-J$ in exactly the same way as the massless boson, but has a different mode expansion for some other linear combination of the quasimomenta.\footnote{Note that in equations $\eqref{f1exp}$ and $\eqref{f3exp}$ that there is a term $\zeta^1f_2^0$ appearing in both $f_1$ and $f_3$. Although these terms cancel when we take the combination $\sin^2\phi\kappa_1+\cos^2\phi\kappa_3$, the presence of $f_2^0$ will produce a different mode expansion for $\kappa_1$ and $\kappa_3$ seperately. In particular, it is important to note that we again have the product of two terms appearing in the expressions for $f_l$. Although the functions $f_l(\sigma)$ are used to write a solution to the generalised residue conditions, it is the actual residues $\kappa_l$ that contain physical information. Upon integrating over $\sigma$, any linear terms in $f_l$, such as the contribution from $\zeta^2$, will have no physical effect, as their contribution can be removed up to a redefinition of the zero modes of the other terms.} This solution is a massless fermion. The quasimomenta that contain both this massless fermion and the massless boson will have residues with $f_2^0$ given by equation $\eqref{fermionmode}$ and $\zeta^1$ containing both mode expansions:
\begin{equation}
\zeta^1(\sigma)=\sqrt{\alpha'}\lambda^{-\frac{1}{4}}(p_0+\psi_0)+\sqrt{2}\lambda^{-\frac{1}{4}}\sum_{n=1}^\infty\sqrt{n}\left((a_n+\psi_n)e^{-in\sigma}+(a_n^\dagger +\psi_n^\dagger)e^{in\sigma}\right)\ .\label{bosandferm}
\end{equation}

The remaining fermion is then generated from the $S^1$ boson in a similar fashion, namely by a (fermionic) mode expansion in $\chi^1$ and $f_2^0$ simultaneously. The full set of massless modes therefore comes from having $\zeta^1$ and $\chi^1$ each with a distinct bosonic and fermionic mode expansion, with both fermionic mode expansions also appearing in $f_2^0$. Each set of excitations contributes identically to $E-J$, but differently for other measurable charges.~\footnote{We would like to thank Olof Ohlsson Sax for a discussion of these issues.} In particular, note that the bosonic massless mode generated from $\chi^1$ is charged under the $U(1)$ charge associated with $S^1$ translations. On the other hand, the mode generated from $\zeta^1$ is neutral under this $U(1)$, so the massless fermion that we generate in the above process from the $S^1$ boson is charged under the $U(1)$ while the fermion generated from the coset boson is not. This difference is natural from the point of view of our finite-gap equations, but is less natural from the point of view of the symmetry algebra of the S-matrix. As such, the representation which the four massless modes form is not obvious from our construction here. The two fermions we derive correspond to two different linear combinations of the fermionic modes which sit naturally within a massless multiplet of the symmetry algebra.

In this sub-section we have used a so-called bosonic grading for the $D(2,1;\alpha)^2$ Cartan algebra used previously in~\cite{Babichenko:2009dk}. In~\cite{Borsato:2012ss} an alternate mixed bosonic-fermionic grading was used to construct the S-matrix of massive excitations. In appendix D we show that at the level of finite gap equations and the GRC the two gradings are equivalent.~\footnote{We would like to thank Alessandro Sfondrini for a discussion of this.}

\subsection{The  BMN limit for $PSU(1,1|2)^2\times (U(1)^4)^2$}

In this subsection we briefly show how the GRC condition applied to $PSU(1,1|2)^2\times (U(1)^4)^2$ finite gap equations can be used to reproduce the BMN limit of the complete (massive and massless) superstring spectrum on $AdS_3\times S^3\times T^4$. Consider first  $AdS_3\times S^3.$ The coset for strings on $AdS_3\times S^3$ is $\frac{PSU(1,1|2)\times PSU(1,1|2)}{SU(1,1)\times SU(2)}$. We take as the Cartan matrix of $PSU(1,1|2)$:
\begin{equation}
A=\left(
\begin{array}{ccc}
&-1&\\-1&2&-1\\&-1&
\end{array}
\right)\ .\label{psu112cartan}
\end{equation}
The quasimomenta for this space are $p_l^\pm$, $l=1,2,3$. The inversion matrix is given by equation $\eqref{d21as}$, and neglecting the windings $m_l^\pm$ for simplicity, we may set $f^+_l=f^-_l\equiv f_l$ . The residue condition $\eqref{d21acon}$ on this coset then reduces to
\begin{equation}
0=\sum_{l,m=1}^3 A_{lm}f_lf_m=2f_2(f_2-f_1-f_3)\ .
\end{equation}
The BMN vacuum has $f_2=0$, and we find that solving the Virasoro condition on the residues implies that $f_2=0$ exactly.~\footnote{The GRC for $AdS_5$ and $AdS_4$ lead to a similar restriction; see the discussion in appendices B and C.} This in turn means there is no contribution from the residues to $E-J$. Hence, as expected, the GRC does not lead to any additional BMN excitations for strings on $AdS_3\times S^3$ alone.

 For strings on $AdS_3\times S^3\times T^4$ we can include the massless modes of $T^4$ much like we  included the massless $S^1$ mode in section 6.2 above. Let us add 4 additional pairs of quasimomenta $p_i^\pm$, $i=1..4$. These have residues $\kappa_i\pm 2\pi m_i$ given in terms of functions $f_i(\sigma)$ just as for the functions $f_l(\sigma)$ giving the residues of the $PSU(1,1|2)$ quasimomenta. With the Cartan matrix for each $U(1)^2$ taken to be the identity and the inversion matrix taken to be $\sigma^1$, the condition $\eqref{d21acon}$ is now
\begin{equation}
0=\sum_{l,m=1}^3 A_{lm}f_l^\pm f_m^\pm+\sum_{i=4}^7(f_i^\pm)^2=2f_2^\pm(f_2^\pm-f_1^\pm-f_3^\pm)+\sum_{i=4}^7(f_i^\pm)^2\ .\label{t4con}
\end{equation}
In fact, we can make an additional simplification in this case. The Cartan matrix $\eqref{psu112cartan}$ has the null eigenvector $(1,0,-1)$. Since it is $A_{lm}\kappa_m$ that appears in the finite-gap equations, we can add the appropriate contributions from any null eigenvector to the residues without changing the finite-gap equations. Therefore we can set $f_1=f_3$. 

The finite-gap equations for the quasimomenta $p_l$ are then given by
\begin{equation}
\pm\frac{\kappa_2z+2\pi m_2}{z^2-1}+2\pi n_{1,i}^\pm=-\dashint\mathrm{d}w\frac{\rho_2^\pm(w)}{z-w}+\dashint\frac{\mathrm{d}w}{w^2}\frac{\rho_2^\mp(w)}{z-\frac{1}{w}}
\end{equation}
\begin{align}
\pm\frac{(\kappa_1-\kappa_2)z+2\pi (m_1-m_2)}{z^2-1}&+2\pi n_{2,i}^\pm=-\dashint\mathrm{d}w\frac{\rho_1^\pm(w)}{z-w}+2\dashint\mathrm{d}w\frac{\rho_2^\pm(w)}{z-w}-\dashint\mathrm{d}w\frac{\rho_3^\pm(w)}{z-w}\nonumber\\
&+\dashint\frac{\mathrm{d}w}{w^2}\frac{\rho_1^\mp(w)}{z-\frac{1}{w}}-2\dashint\frac{\mathrm{d}w}{w^2}\frac{\rho_2^\mp(w)}{z-\frac{1}{w}}+\dashint\frac{\mathrm{d}w}{w^2}\frac{\rho_3^\mp(w)}{z-\frac{1}{w}}
\end{align}
\begin{equation}
\pm\frac{\kappa_2z+2\pi m_2}{z^2-1}+2\pi n_{3,i}^\pm=-\dashint\mathrm{d}w\frac{\rho_2^\pm(w)}{z-w}+\dashint\frac{\mathrm{d}w}{w^2}\frac{\rho_2^\mp(w)}{z-\frac{1}{w}}
\end{equation}
which should be taken together with the fact that the residues are given in terms of the functions $f_l$ via equation $\eqref{c1}$ and these functions satisfy equation $\eqref{t4con}$. The quasimomenta $p_i$ associated to the $T^4$ directions trivially satisfy their own finite-gap equations with no cuts.

Now we will derive the massless components of the BMN spectrum using the generalised residue conditions. $p_2$ is the only quasimomentum associated to a momentum carrying node in the Dynkin diagram, and so $f_2$ is the only function that contributes to $E-J$. We can solve equation $\eqref{t4con}$ to give all other functions in terms of $f_1$ and 4 new functions $\zeta_i$, $i=4\dots 7$. Taking $f_3=f_1$ as above and neglecting winding so we rewrite equation $\eqref{t4con}$ as
\begin{equation}
2(f_2-f_1)^2+\sum_{i=4}^7f_i^2=2f_1^2
\end{equation}
The solution to this can be given by
\begin{align}
&f_4=\sqrt{2}f_1\sin\zeta_4\nonumber\\
&f_5=\sqrt{2}f_1\cos\zeta_4\sin\zeta_5\nonumber\\
&f_6=\sqrt{2}f_1\cos\zeta_4\cos\zeta_5\sin\zeta_6\nonumber\\
&f_7=\sqrt{2}f_1\cos\zeta_4\cos\zeta_5\cos\zeta_6\sin\zeta_7\nonumber\\
&f_2=f_1(1-\cos\zeta_4\cos\zeta_5\cos\zeta_6\cos\zeta_7)\ .
\end{align}
For the BMN vacuum we have $f_1=f_3=\kappa$ and $f_2=0$, and expanding the residues at large $\kappa$ we find  $\zeta_i=0$ and hence $f_i=0$ for $i=4..7$. Therefore, the large $\kappa$ expansions are
\begin{equation}
f_1(\sigma,\kappa)=\kappa+f_1^0(\sigma)+\frac{1}{\kappa}f_1^1(\sigma)+\mathcal{O}\left(\frac{1}{\kappa^2}\right), \quad \zeta_i(\sigma,\kappa)=\frac{1}{\kappa}\zeta_i^1(\sigma)+\mathcal{O}\left(\frac{1}{\kappa^2}\right)\ ,
\end{equation}
and we have
\begin{equation}
E-J\sim\int_0^{2\pi}\mathrm{d}\sigma f_2(\sigma)=\frac{1}{2\kappa}\sum_{i=4}^7\int_0^{2\pi}\mathrm{d}\sigma \zeta_i^1(\sigma)^2 +\mathcal{O}\left(\frac{1}{\kappa^2}\right)\ .
\end{equation}
We have four integrals of the squares of periodic functions over their periods, giving four mode expansions contributing to $E-J$ at $\mathcal{O}\left(\frac{1}{\kappa}\right)$, just as we expect for the four massless bosonic modes.

The massless fermions are generated from the massless bosons in a way similar to what was done in section 6.2, namely by making fermionic mode expansions in $\zeta_i^1(\sigma)$ and $f_1^1(\sigma)$ simultaneously. The full massless spectrum therefore comes from each $\zeta_i^1$ containing both a bosonic and fermionic mode expansions, as in equation $\eqref{bosandferm}$, while $f_1^1$ contains all four of these fermionic mode expansions. The massive spectrum analysis follows from~\cite{Babichenko:2009dk}.

\section{Conclusion}

In this paper we have re-examined the derivation of finite-gap equations for string theories on semi-symmetric cosets. These equations govern the analytic properties of quasi-momenta $p_l(z)$. The quasi-momenta can have cuts and simple poles in the complex $z$ plane. In section 3.2 we found that the residue condition~$\eqref{nullres}$\footnote{Equation \eqref{cartanoldcon} for a non-trivial Cartan matrix.} used in the previous literature is stronger than the one required by the Virasoro constraints. Instead, we showed that the conditions implied by the Virasoro constraints are the more general ones $\eqref{c1}$ and $\eqref{c2}$\footnote{Equation \eqref{d21acon} for a non-trivial Cartan matrix.} the second of which we have called the generalised residue condition. In section 4 we considered classical string solutions on $\mathbb{R}\times S^1\times S^1$ in order to demonstrate explicitly how the Virasoro constraints are equivalent to the generalised residue conditions but not the null condition $\eqref{nullres}$.~\footnote{In Appendix A we show this same result for $\mathbb{R}\times S^3\times S^1$, and it is clear from there to see why it is true for the full geometry, or indeed other backgrounds.} When we studied explicit classical solutions containing massless excitations, we saw that the residues of their quasimomenta did not satisfy the condition $\eqref{nullres}$, and so relaxing this condition to $\eqref{c1}$ and $\eqref{c2}$ was necessary to derive the massless mode from the finite-gap equations. Then in sections 5 and 6 we saw that this was also sufficient; taking the GRC it is possible to derive the complete spectrum in the BMN limit of the finite-gap equations.

It might seem surprising that the method used to determine the massless modes should be somewhat different from the method used to determine the two lightest massive modes, leading us to wonder if there exists a more concise procedure that can be applied to all the modes. However, from the explicit quasimomenta we constructed in section 4, we can see why this distinct approach is in fact necessary. The quasimomenta of these explicit solutions did not contain any branch cuts, in contrast to any quasimomenta containing a massive excitation. The BMN limit manifests itself at the level of the quasimomenta as a limit in which the cuts shrink to a set of isolated points, and the massive modes are found by considering the finite-gap equations in that limit. For solutions with no cuts, such as the quasimomenta in section 4, there are technically no finite-gap equations. We suggest that the correct way to regard these apparently different methods consistently is to add an additional notion to the interpretation of the BMN limit from the perspective of the quasimomenta. As well as taking a limit where the cuts shrink, the BMN limit also involves taking a limit of the residues towards their BMN vacuum values. 

Finite-gap equations have been written down for string theory on other cosets, notably those corresponding to the backgrounds $AdS_5\times S^5$ and $AdS_4\times CP^4$. In these backgrounds, the full BMN spectra can be derived from the finite-gap equations without the need to generalise the residue condition $\eqref{nullres}$ to $\eqref{c1}$ and $\eqref{c2}$. In Appendices B and C we give the results of applying the generalised residue analysis to these backgrounds, to show that there are no additional BMN modes produced by the generalised residues in these cases. On more general cosets however, the GRC may lead to non-trivial corrections to the residue conditions used in the literature. For example we expect such effects to arise in the $AdS_2\times S^2\times S^2\times T^4$ 
theories~\cite{ Sorokin:2011rr,Cagnazzo:2011at,Murugan:2012mf,Cagnazzo:2012uq,Abbott:2013kka}.

It would be interesting to see how the GRC conditions appear from the thermodynamic limit of the Bethe Ansatz and whether they can help to resolve some of the discrepancies observed in~\cite{Borsato:2012ss}.~\footnote{We would like to thank Riccardo Borsato and Alessandro Sfondrini for discussions about this.} Another potentially interesting question is whether one could understand how to incorporate the massless modes into the Landau-Lifshitz sigma models that encode the large-charge limit of the string sigma model~\cite{Kruczenski:2003gt,Kruczenski:2004kw,Hernandez:2004uw,
Stefanski:2004cw,Stefanski:2005tr,Stefanski:2007dp}

\section*{Acknowledgements}

We would like to thank Kostya Zarembo for many discussions and for insightful comments at various stages of this work. We would also like to thank R. Borsato, N. Gromov, O. Ohlsson Sax and A. Sfrondrini for helpful discussions. B.S. is grateful to CERN and to the Centro de Ciencias de Benasque Pedro Pascual in Benasque for hospitality during parts of this work. T.L. is supported by an STFC studentship. We acknowledge partial support under the STFC Consolidated Grant ``Theoretical Physics at City University” ST/J00037X/1.

\appendix
\section{Residues of quasimomenta on $\mathbb{R}\times S^3\times S^1$}
The metric is
\begin{equation}
ds^2=R^2\left[-dt^2+\frac{1}{\cos^2\phi}(d\theta^2+\cos^2\theta d\psi_1^2+\sin^2\theta d\varphi^2)+\frac{1}{\sin^2\phi}d\psi_2^2\right]\ . \label{EE}
\end{equation}
The group representative $g$ is a direct sum $g=g_0\oplus g_1\oplus g_2$ as before. $g_0$ and $g_2$ are chosen exactly as in $\eqref{g0}$ and $\eqref{g1g2}$, but for  $g_1$ corresponding to the full $S^3$ we take
\begin{equation}
g_1=\sqrt{\frac{1}{2\cos\phi}}\left(
\begin{array}{cccc}
\cos\theta e^{i\psi_1} & -\sin\theta e^{-i\varphi} & 0 & 0\\
\sin\theta e^{i\varphi} & \cos\theta e^{-i\psi_1} & 0 & 0\\
0 & 0 &  i\sin\theta e^{-i\varphi} & -i\cos\theta e^{i\psi_1}\\
0 & 0 & i\cos\theta e^{-i\psi_1} & -i\sin\theta e^{i\varphi}
\end{array}\right)\ .
\end{equation}
The current $j$ is (with the first and third terms in the direct sum unchanged from equation $\eqref{jmassless}$)
\begin{align}
j=\frac{dt}{2}\left(
\begin{array}{cccc}
0&1&0&0\\1&0&0&0\\0&0&0&-1\\0&0&-1&0
\end{array}
\right)&\oplus 
\frac{1}{2\cos\phi}\left(
\begin{array}{cccc}
iu & -v+iw & 0 & 0\\
v+iw & -iu & 0 & 0\\
0 & 0 & iu & -v-iw\\
0 & 0 & v-iw & -iu
\end{array}
\right)\nonumber\\
&\oplus
\frac{i}{\sin\phi}\frac{d\psi_2}{2}\left(
\begin{array}{cccc}
1&0&0&0\\0&-1&0&0\\0&0&1&0\\0&0&0&-1
\end{array}
\right)\ ,\label{js3}
\end{align}
where $u$, $v$ and $w$ are all real one-forms given by
\begin{align}
&u=\cos^2\theta d\psi_1+\sin^2\theta d\varphi\nonumber\\
&v+iw=e^{i(\psi_1+\varphi)}\left(d\theta +i\sin\theta\cos\theta(d\varphi-d\psi_1)\right)\ . \label{uvw}
\end{align}
As in section 4, we have again chosen a group representative satisfying $\Omega(j)=-j$ and so $j^{(2)}=\frac{1}{2}(j-\Omega(j))=j$. We can confirm that
\begin{align}
\mathrm{tr}\left[(j^{(2)})^2\right]=\mathrm{tr}(j^2)&=dt^2-\frac{1}{\cos^2\phi}\left(u^2+v^2+w^2\right)-\frac{1}{\sin^2\phi}d\psi_2^2\nonumber\\&=dt^2-\frac{1}{\cos^2\phi}\left(d\theta^2+\cos^2\theta d\psi_1^2+\sin^2\theta d\varphi^2\right)-\frac{1}{\sin^2\phi}d\psi_2^2\ .\label{s3metricuvw}
\end{align}

The relevant ($S^3$) part of the Lax operator $L_\sigma$ obtained from the current in $\eqref{js3}$ is given by
\begin{equation}
L_\sigma=\left(
\begin{array}{cccc}
ia&-b+ic&0&0\\
b+ic&-ia&0&0\\
0&0&ia&-b-ic\\
0&0&b-ic&-ia
\end{array}\right)\ , \label{s3lax}
\end{equation}
with $a$, $b$ and $c$ given by
\begin{align}
&a=\frac{1}{2\cos\phi}\frac{1}{z^2-1}\left[(z^2+1)u_\sigma+2zu_\tau\right]\ ,\nonumber\\
&b=\frac{1}{2\cos\phi}\frac{1}{z^2-1}\left[(z^2+1)v_\sigma+2zv_\tau\right]\ ,\nonumber\\
&c=\frac{1}{2\cos\phi}\frac{1}{z^2-1}\left[(z^2+1)w_\sigma+2zw_\tau\right]\ .
\end{align}

We can find the residues of the quasimomenta on this space using the WKB analysis (see section 3.2). We need the eigenvalues of $V=-ihL_\sigma$ in the limit $h=z\mp1\to 0$. With $L_\sigma$ as in equation $\eqref{s3lax}$, there is the following eigenvalue of multiplicity 2:
\begin{equation}
\frac{1}{2\cos\phi}\sqrt{(u_\tau\pm u_\sigma)^2+(v_\tau\pm v_\sigma)^2+(w_\tau\pm w_\sigma)^2}
\end{equation}
and of course the negative of this. Note that $\pm$ in this expression refers to the limit $z\to\pm 1$. 

We therefore have expressions for the residues of the quasimomenta on this space as follows. There are residues $\kappa_0\pm 2\pi m_0$ and $\kappa_3\pm 2\pi m_2$ given as in equation $\eqref{kappasms}$ for the quasimomenta associated to $\mathbb{R}$ and $S^1$. There are generically two distinct quasimomenta $p_1^+$ and $p_1^-$ associated to $S^3$, but they both have the same residues (with opposite signs as required by the inversion symmetry); this equality of residues is seen in the fact that the residues of $V$ have multiplicity two. These residues are
\begin{equation}
\kappa_1\pm 2\pi m_1=\frac{1}{\cos\phi}\int_0^{2\pi}\mathrm{d}\sigma\sqrt{(u_\tau\pm u_\sigma)^2+(v_\tau\pm v_\sigma)^2+(w_\tau\pm w_\sigma)^2}\ .
\end{equation}
We can therefore see that the residues for all quasimomenta, including those on $S^3$, are given naturally in terms of integrals of functions $f_l^\pm(\sigma)$. Furthermore, using equation $\eqref{s3metricuvw}$, we can see that the condition $\eqref{c2}$ on these functions is exactly the more familiar form of the Virasoro constraints on classical bosonic strings on a curved background, here $\mathbb{R}\times S^3\times S^1$, namely
\begin{equation}
G_{\mu\nu}(\dot{X}^\mu\pm X'^\mu)(\dot{X}^\nu\pm X'^\nu)=0\label{virspace}
\end{equation}
where $X^\mu$ are the spacetime fields and $G_{\mu\nu}$ is the spacetime metric.

Similarly for the quasimomenta for the full coset space of $AdS_3\times S^3\times S^3$, the Virasoro constraints in the form $\eqref{virspace}$ can be seen to be equivalent to the generalised residue conditions $\eqref{c1}$ and $\eqref{c2}$, not the null residue condition $\eqref{nullres}$.

\section{Generalised residue conditions for $AdS_5\times S^5$}
The coset for strings on $AdS_5\times S^5$ is $\frac{PSU(2,2|4)}{SO(4,1)\times SO(5)}$. We follow the conventions of the review \cite{Zarembo:2010yz}. The Cartan matrix for $PSU(2,2|4)$ is
\begin{equation}
A=\left(
\begin{array}{ccccccc}
&1&&&&&\\
1&-2&1&&&&\\
&1&&-1&&&\\
&&-1&2&-1&&\\
&&&-1&&1&\\
&&&&1&-2&1\\
&&&&&1&
\end{array}
\right)
\end{equation}
and the matrix $S$ giving the inversion symmetry through equation $\eqref{inversion}$ is
\begin{equation}
S=\left(
\begin{array}{ccccccc}
&&1&-1&&&\\
&1&&-1&&&\\
1&&&-1&&&\\
&&&-1&&&\\
&&&-1&&&1\\
&&&-1&&1&\\
&&&-1&1&&
\end{array}
\right)\ .
\end{equation}

The quasimomenta are $p_l$ with the index $l$ running from 1 to 7. The residues are given in terms of functions $f_l(\sigma)$ as in equation $\eqref{c1}$. The action of the inversion symmetry on the residues (see equation $\eqref{symres}$) means $f_l$ must satisfy
\begin{equation}
\sum_{m=1}^7 S_{lm}f_m=-f_l\ .
\end{equation}
Solving this inversion symmetry, we find that we can choose $f_1$, $f_4$ and $f_7$ to be independent, while the remaining functions are given in terms of these three:
\begin{equation}
f_2=f_6=\frac{1}{2}f_4, \quad f_3=f_4-f_1, \quad f_5=f_4-f_7\ .
\end{equation}
With these substitutions made, the version of the condition $\eqref{d21acon}$ on this space is
\begin{equation}
0=\sum_{l,m=1}^7 A_{lm}f_lf_m=f_4\left(f_1+f_7-\frac{1}{2}f_4\right)\ .\label{ads5con}
\end{equation}

The values of $f_l$ for the BMN vacuum are
\begin{equation}
f_1+f_7=\kappa,\quad f_4=0\ .
\end{equation}
For the residues of $D(2,1;\alpha)^2$ we were able to solve the constraint on the functions $f_l$ in a way that allowed an expansion around the BMN vacuum. Here however, we can see that there is no way to solve the condition $\eqref{ads5con}$ in any other way than setting $f_4=0$ when we take a similar approach. Suppose we make an expansion in large $\kappa$ as follows:
\begin{equation}
f_4(\sigma,\kappa)=f_4^0(\sigma)+\frac{1}{\kappa}f_4^1(\sigma)+\mathcal{O}\left(\frac{1}{\kappa^2}\right), \quad f_1+f_7=\kappa+f_1^0+f_7^0+\frac{1}{\kappa}(f_1^1+f_7^1)+\mathcal{O}\left(\frac{1}{\kappa^2}\right)\ .
\end{equation}
Then we can insert these equation $\eqref{ads5con}$ and require that it holds order by order. At $\mathcal{O}(\kappa)$ we require $f_4^0=0$. Then, using this together with the requirement that equation $\eqref{ads5con}$ holds at $\mathcal{O}(1)$ we require $f_4^1=0$ and so on. If we assume that this perturbative expansion around the BMN vacuum gives us every possible state, then we conclude that we must have $f_4=0$ identically. This reproduces the usual finite-gap equations for this space. In addition $p_4$ corresponds to the only mode in the Dynkin diagram which carries energy and momentum, and $E-J$ is given solely in terms of $p_4$. The fact that $f_4=0$, and hence $\kappa_4=0$, means that there is no contribution to $E-J$ from the residues.

\section{Generalised residue conditions for $AdS_4\times CP^4$}
The coset for strings on $AdS_4\times CP^4$ is $\frac{OSp(6|4)}{U(3)\times SO(3,1)}$. The Cartan matrix of $OSp(6|4)$ is
\begin{equation}
A=\left(
\begin{array}{ccccc}
&1&&&\\
1&-2&1&&\\
&1&&-1&-1\\
&&-1&2&\\
&&-1&&2
\end{array}
\right)
\end{equation}
and the inversion symmetry matrix $S$ is
\begin{equation}
S=\left(
\begin{array}{ccccc}
&&1&-1&-1\\
&1&&-1&-1\\
1&&&-1&-1\\
&&&&-1\\
&&&-1&
\end{array}
\right)\ .
\end{equation}
Now the quasimomenta are $p_l$ with $l$ running from 1 to 5. The action of the inversion symmetry on the residues means that there are 2 independent functions $f_1$ and $f_4$, with the others given by
\begin{equation}
f_2=f_5=f_4, \quad f_3=2f_4-f_1\ .
\end{equation}
Then in terms of $f_1$ and $f_4$, the condition the functions need to satisfy is
\begin{equation}
0=\sum_{l,m=1}^5 A_{lm}f_lf_m=2f_4(2f_1-f_4)\ .
\end{equation}
We see that this is very similar in form to the condition $\eqref{ads5con}$, and the argument from this point is identical to that in the last section. The BMN vacuum has $f_4=0$ and $f_1=\kappa$, and expanding around the BMN vacuum we find there is no way to add non-zero terms to $f_4$. The contributions to $E-J$ in this space come only from $p_4$ and $p_5$, and we noted that $f_5=f_4$. Hence there is no contribution to $E-J$ from the residues.

\section{$D(2,1;\alpha)^2/SU(1,1)\times SU(2)^2$ in mixed grading}

In section 6, we used a grading for $D(2,1;\alpha)^2$ which involves bosonic Cartan generators only. In \cite{Borsato:2012ss} an alternative grading was used, involving bosonic Cartan generators on one factor of $D(2,1;\alpha)$ and fermionic generators on the other. The Cartan matrix is given in this mixed grading by
\begin{equation}
A=\left(
\begin{array}{cccccc}
4\sin^2\phi&-2\sin^2\phi&&&&\\
-2\sin^2\phi&&-2\cos^2\phi&&&\\
&-2\cos^2\phi&4\cos^2\phi&&&\\
&&&&2\sin^2\phi&-2\\
&&&2\sin^2\phi&&2\cos^2\phi\\
&&&-2&2\cos^2\phi&
\end{array}
\right)
\end{equation}
and the matrix $S$ defining the action of the inversion symmetry on the quasimomenta through equation $\eqref{inversion}$ is given by
\begin{equation}
S=\left(
\begin{array}{ccc}
-1&&\\-1&1&-1\\&&-1
\end{array}
\right)
\otimes \sigma_1\ .
\end{equation}
Following the notation in \cite{Borsato:2012ss}, we take the index structure on the quasimomenta as follows: we have quasimomenta $p_l$ and $p_{\bar{l}}$ with $l,\bar{l}=1,2,3$. The upper left quadrant of $A$ corresponds to indies $l$, the lower right to indices $\bar{l}$, and the factor of $\sigma_1$ in $S$ interchanges $l$ and $\bar{l}$.

The action of the inversion symmetry on the residues via equation $\eqref{symres}$ means we can determine the functions $f_{\bar{l}}$ in terms of $f_l$. We have:
\begin{equation}
f_{\bar{1}}=f_1, \quad f_{\bar{3}}=f_3, \quad f_{\bar{2}}=f_1-f_2+f_3\ .
\end{equation}
We can insert this into the relevant equivalent of the condition $\eqref{d21acon}$ and we find that:\footnote{$A_{lm}$ referring only to the upper-left components of $A$ and $A_{\bar{l}\bar{m}}$ to the lower-right components.}
\begin{equation}
\sum_{l,m}^3A_{lm}f_lf_m=\sum_{\bar{l},\bar{m}=1}^3A_{\bar{l}\bar{m}}f_{\bar{l}}f_{\bar{m}}=4\sin^2\phi f_1(f_1-f_2)+4\cos^2\phi f_3(f_3-f_2)\ .
\end{equation}
In other words, in the mixed grading just as in the bosonic grading, the residue condition is identical when considered either solely on left-movers or right-movers. The full condition in this case is
\begin{equation}
\sum_{l,m}^3A_{lm}f_lf_m+\sum_{\bar{l},\bar{m}=1}^3A_{\bar{l}\bar{m}}f_{\bar{l}}f_{\bar{m}}=0
\end{equation}
and so we have exactly the same condition with exactly the same analysis for quasimomenta in the mixed grading as in bosonic grading.

\end{document}